\documentclass[aps,pra,amsmath,amssymb,twocolumn,superscriptaddress,notitlepage]{revtex4-2}
\usepackage{graphicx}
\usepackage{color}
\usepackage{tikz}
\usepackage{braket}
\usepackage{titlesec}
\usepackage{bm}
\usepackage{xcolor}
\usepackage{framed}
\usepackage[linesnumbered,ruled,vlined]{algorithm2e}
\usepackage[colorlinks,citecolor=blue,linkcolor=red,urlcolor=blue]{hyperref}
\hypersetup{
  pdfkeywords={},
  pdftitle={Quantum Cram\'er-Rao Precision Limit of Noisy Continuous Sensing},
  }

\makeatletter

\definecolor{DYblue}{RGB}{1,100,0}

\colorlet{shadecolor}{blue!10}

\titleformat{\section}
  {\centering\bfseries}
  {}                             % a prefix
  {0pt}                          % How much space exists between the prefix and the title
  {APPENDIX \thesection:\quad\MakeTextUppercase}    % How the section is represented

\begin{document}

\title{Quantum Cram\'er-Rao Precision Limit of Noisy Continuous Sensing}
\author{Dayou Yang}
\email{dayou.yang@ustc.edu.cn}
\affiliation{Hefei National Laboratory, University of Science and Technology of China, Hefei 230088, China}
\affiliation{Shanghai Research Center for Quantum Sciences and CAS Center for Excellence in Quantum Information and Quantum Physics, University of Science and Technology of China, Shanghai 201315, China}
\affiliation{
Institut f\"ur Theoretische Physik and IQST, Universit\"at Ulm,
Albert-Einstein-Allee 11, D-89069 Ulm, Germany}

\author{Moulik Ketkar}
\affiliation{
Institut f\"ur Theoretische Physik and IQST, Universit\"at Ulm,
Albert-Einstein-Allee 11, D-89069 Ulm, Germany}
\affiliation{
Department of Physics, Indian Institute of Technology (BHU),
Varanasi 221005, India}
\author{Koenraad Audenaert}
\affiliation{
Institut f\"ur Theoretische Physik and IQST, Universit\"at Ulm,
Albert-Einstein-Allee 11, D-89069 Ulm, Germany}
\author{Susana F. Huelga}
\email{susana.huelga@uni-ulm.de}
\affiliation{
Institut f\"ur Theoretische Physik and IQST, Universit\"at Ulm,
Albert-Einstein-Allee 11, D-89069 Ulm, Germany}
\author{Martin B. Plenio}
\email{martin.plenio@uni-ulm.de}
\affiliation{
Institut f\"ur Theoretische Physik and IQST, Universit\"at Ulm,
Albert-Einstein-Allee 11, D-89069 Ulm, Germany}

\date{\today}

\begin{abstract}
Quantum sensors hold considerable promise for precision measurement, yet their capabilities are inherently constrained by environmental noise. A fundamental task in quantum sensing is determining the precision limit of noisy sensor devices. For continuously monitored quantum sensors, characterizing the optimal precision in the presence of environments other than the measurement channel is an outstanding open theoretical challenge, due to the infinite-dimensional nature of the sensor output field and the complex temporal correlation of the photons therein. Here, we establish a numerically efficient method to determine the quantum Cram\'er-Rao bound for continuously monitored quantum sensors subject to general environmental noise---Markovian or non-Markovian, and showcase its application with paradigmatic models of continuously monitored quantum sensors. Applicable to both constant-parameter and waveform estimation, our method provides a rigorous and practical framework for assessing and enhancing the sensor performance in realistic settings, with broad applications across experimental quantum physics.
\end{abstract}
\maketitle
The performance of quantum sensors is fundamentally limited by experimental noise and imperfections; in extreme cases, even minimal noise can negate the anticipated quantum-enhanced precision~\cite{PhysRevLett.79.3865}. A central challenge in quantum sensing is thus to characterize precision limit of sensors operating under noise. Research typically 
addresses this by establishing precision bounds applicable to various sensing scenarios~\cite{Escher:2011aa,Demkowicz-Dobrzanski:2012aa,PhysRevLett.116.120801,PhysRevX.7.041009}, though these bounds may not be guaranteed to be attainable. An alternative, often more practical, approach is to develop methods to evaluate the optimally \emph{attainable} precision for specific (classes of) sensor setups and noise models~\cite{PhysRevA.84.012103,PhysRevLett.109.233601,Chabuda:2020tv,PhysRevLett.127.060501}. This approach establishes stringent criteria for evaluating whether a sensor design meets the precision requirements of a target application, and guide experimental development towards optimal performance by suggesting improved design and parameter choices. A widely adopted precision limit in this context is the quantum Cram\'er-Rao bound (QCRB)~\cite{Helstrom1976}. For single-parameter estimation, it specifies the optimally attainable mean squared error (MSE); while for multi-parameter estimation, e.g., waveform estimation~\cite{PhysRevLett.106.090401,PhysRevLett.111.113601,PhysRevA.97.042334}, it provides a rigorous lower bound on the MSE. For unbiased estimation, the QCRB is expressed via the inverse of the quantum Fisher information (QFI)~\cite{PhysRevLett.72.3439}. 
\begin{figure}[b!]
\centering{} \includegraphics[width=0.48\textwidth]{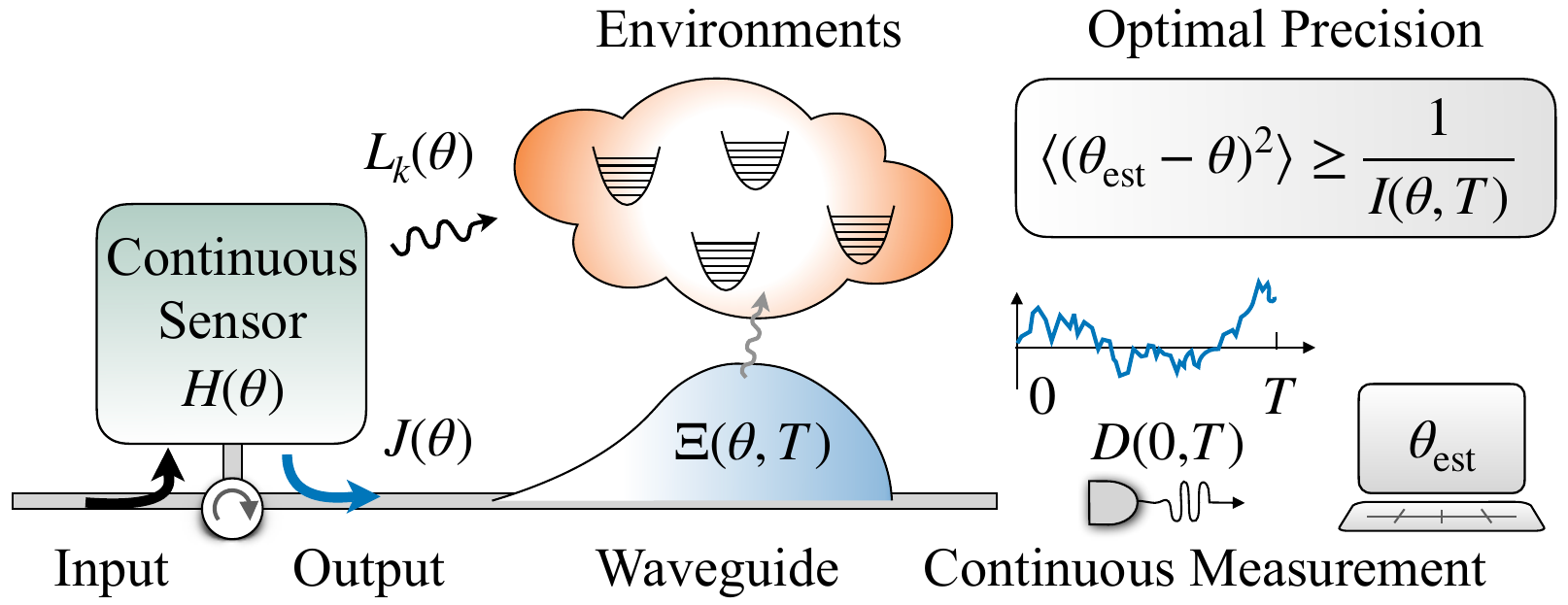} 
\caption{Noisy continuous sensors and their precision limit. The sensor is an open quantum-optical device whose dynamics {depend} on the unknown parameter $\theta$. The sensor is driven by an input quantum field (e.g., waveguide photons), with its output field subject to continuous measurement spanning a time window $[0,T]$, yielding a record $D(0,T)$ from which $\theta$ can be estimated. The optimal precision of unbiased estimation is quantified by the quantum Fisher information (QFI) $I(\theta,T)$ of the output-field quantum state $\Xi(\theta,T)$. The sensor and waveguide may couple with surrounding environments, resulting in decoherence that fundamentally constrains $I(\theta,T)$. A generic noisy continuous sensor is characterized by its Hamiltonian $H(\theta)$ and jump operator $J(\theta)$ $[L_k(\theta)]$ responsible for the sensor-waveguide [sensor-environment] coupling. We establish a unified framework for evaluating the QFI for arbitrarily driven continuous sensors subjected to general (including non-Markovian) environmental noise, {which further accommodates waveform estimation scenarios where the constant-parameter $\theta$ is replaced by a function $\vartheta(t),t\in[0,T]$}.}
\label{fig:fig1} 
\end{figure}

A broad class of important sensor technology relies on continuous measurement~\cite{Braginsky1992,carmichael1993open,RevModPhys.70.101,gardiner2004quantum,Wiseman,gardiner2015quantum}, wherein an open quantum sensor is driven by an input quantum field (e.g., a laser) and continuously monitored through detection of the output field, as illustrated in Fig.~\ref{fig:fig1}. Prominent examples range from gravitational-wave detectors~\cite{PhysRevLett.116.061102,Danilishin:2012wp}, via atomic gas magnetometers~\cite{Moller:2017ta,Colangelo:2017wi}, to driven-dissipative interacting many-body sensors~\cite{PRXQuantum.3.010354,Ding:2022aa,doi:10.1126/sciadv.ado8130,Ilias:2024aa,PhysRevLett.132.050801}. These setups are inherently noisy, as both the open sensor and the quantum field may decohere via coupling with surrounding environments---e.g., in the form of dephasing, emission into inaccessible modes or field attenuation. As only the designated output field is detected, the QCRB of continuous sensors can be rigorously quantified by the QFI of the monitored output field {\em in the presence} of environmental noise. However, the photons of the output field populate infinitely many photonic modes and possess intricate correlation both among themselves and with the external environments, posing an outstanding challenge for QCRB evaluation. Although loose precision bound may be obtained from judicious purification~\cite{Tsang_2013}, evaluating the QFI of the output field plus all environments~\cite{PhysRevX.13.031012} or additionally including the sensor~\cite{PhysRevLett.106.090401,PhysRevLett.112.170401,2015JPhA...48J5301C}, they do not provide stringent calibration of the sensor performance under noise and hence offer limited utility for practical applications.

In this Letter, we establish a numerically efficient method for evaluating the QCRB of generic continuously monitored quantum sensors subject to arbitrary (including non-Markovian) environmental noise. Our method applies to arbitrary sensor dynamics, including non-Gaussian and time-dependent cases, with any input quantum field. {Furthermore, it naturally extends to evaluating the QCRB for noisy waveform estimation~\cite{PhysRevLett.106.090401,PhysRevLett.111.113601,PhysRevA.97.042334}.}

The key insight underlying our method is that, although the output field formally contains infinitely many photonic frequency modes, its quantum state can be efficiently represented as continuous matrix product operators (cMPOs)~\cite{RevModPhys.93.045003} in the infinitesimal time-bin basis. The structure of this cMPO is determined entirely by the open dynamics of the sensor. {This correspondence, remarkably, allows for the efficient evaluation of the output-field QFI by formulating and solving dynamical equations solely in terms of (replicas of) the open sensor variables, eliminating the need for explicit numerical discretization and manipulation of the output-field cMPO}—a process that typically requires low excitation density and may suffer from numerical instability~\cite{PhysRevLett.109.020604}. As such, our approach offers a similar level of clarity as methods~\cite{PhysRevLett.112.170401,PhysRevX.13.031012} developed for the noiseless scenario.
We term our approach \emph{generalized replica master equations} (GRMEs) and develop numerically efficient method for solving them. Beyond quantum sensing, the GRMEs established here extend to evaluating general quantum informational properties that depend nonlinearly on the output-field density operator, which we address in a follow-up article~\cite{Dayou}.

\emph{Quantum Optical Model.---}We model the sensor as a quantum-optical open system coupled with a unidirectional one-dimensional photonic field (hereafter referred to as the waveguide) as its input-output channel, with associated annihilation [creation] operators $f(\omega)$ $[f^\dag(\omega)]$ for photons at frequency $\omega$. Following standard quantum-optical treatment~\cite{carmichael1993open,RevModPhys.70.101,breuer2002theory,gardiner2004quantum,Wiseman,gardiner2015quantum,Rivas_2012}, we consider photons in a bandwidth $\cal B$ around the mean optical frequency $\bar{\omega}$, and introduce the associated quantum noise operator ($\hbar=1$ hereafter)
\begin{equation}
f(t)=\frac{1}{\sqrt{2\pi}}\int_{\cal B} d\omega f(\omega)e^{-i(\omega-\bar{\omega})t}\nonumber
\end{equation}
satisfying the white noise commutation relation $[f(t),f^\dag(t')]=\delta(t-t')$. Decoherence sources can be modeled conveniently via factorization into a series of unidirectional one-dimensional bosonic environments interacting with the sensor~\footnote{This modeling can accommodate both sensor decoherence and waveguide decoherence, with the latter incorporated via suitable global unitary transformations involving the sensor and the waveguide.}. We temporarily assume the environments are Markovian, and later extend our analysis to the non-Markovian case. We can therefore introduce quantum noise operators $b_k(t),b_k^\dag(t)$ for each environment $k=1,2,\dots,K$ with the white noise condition $[b_k(t),b_j^\dag(t^\prime)]=\delta_{kj}\delta(t-t')$. Aiming to sense a general parameter $\theta$, the global evolution of the sensor, waveguide and all environments in the interaction picture is hence described by a Stratonovich quantum stochastic Schr\"odinger equation~\cite{carmichael1993open,RevModPhys.70.101,gardiner2004quantum,Wiseman}, $i(d/dt)\ket{\Psi}=H_{\rm Glob}\ket{\Psi}$, with
\begin{equation}
\label{eq:HSE}
H_{\rm Glob}=H(\theta)+i\left[J(\theta) f^{\dag}(t)+\sum_{k} L_k(\theta)b_k^\dag(t)-{\rm h.c.}\right].
\end{equation}
Here $H$ is the sensor Hamiltonian, $J$ and $L_k$ are the sensor jump operator (with jump rate absorbed into their definitions) that couple to the waveguide and the $k$-th environment respectively, and h.c. denotes the Hermitian conjugate. The extension to sensing a waveform $\vartheta(t)$ is straightforward, {see End Matter}. 

The quantum evolution according to Eq.~\eqref{eq:HSE} formally corresponds to the canonical scheme for generating continuous matrix-product states (cMPSs)~\cite{PhysRevLett.104.190405,PhysRevLett.105.260401}. To illustrate this, let us assume for simplicity factorizable initial state with both the waveguide and environments in vacuum, $\ket{\Psi(0)}=\ket{\psi_{\rm S}(0)}\otimes\ket{\rm vac}_{\rm W,E}$ (hereafter the subscripts S,~W,~E represents the sensor, waveguide and environments respectively). This allows us to transform Eq.~\eqref{eq:HSE} from the Stratonovich into the It\^o form~\cite{carmichael1993open,RevModPhys.70.101,gardiner2004quantum,Wiseman,gardiner2015quantum} and to formally write down its solution as a multi-field cMPS~\cite{RevModPhys.93.045003,PhysRevLett.105.260401,PhysRevLett.104.190405}
\begin{align}
\ket{\Psi(\theta, T)}=&{\cal T}{\rm exp}\bigg\{\int_0^T dt[Q_{\rm}(\theta)+J(\theta) f^\dag(t)\nonumber\\
&+\sum_k L_k(\theta) b_k^\dag(t)]\bigg\}\ket{\psi_{\rm S}(0)}\otimes\ket{\rm vac}_{\rm W,E}\nonumber
\end{align}
with ${\cal T}$ denoting time-ordering and $Q:=-iH-(J^\dag J+\sum_k L_k^\dag L_k)/2$. Tracing away the waveguide and environments provides the reduced density operator of the sensor, $\rho(\theta,T):={\rm tr_{W,E}}[\ket{\Psi(\theta,T)}\bra{\Psi(\theta,T)}]$, whose evolution follows a Lindblad master equation (LME)
\begin{subequations}
\begin{align}
\dot{\rho}=&J(\theta)\rho J^\dag(\theta)+{\cal L}_{0}(\theta)\rho,\label{eq:LME_sys}\\
{\cal L}_{\rm 0}(\theta)\rho:=&-i[H(\theta),\rho]-\frac{1}{2}\{J^\dag(\theta) J(\theta),\rho\}\nonumber
\\&+\sum_{k}{\cal D}[L_k(\theta)]\rho,\label{eq:no_jump_evol}
\end{align}
\end{subequations}
wherein $\{\cdot,\cdot\}$ is the anti-commutator and ${\cal D}[L]\rho:= L\rho L^\dag-\frac{1}{2}\{L^\dag L,\rho\}$ denotes the Lindblad superoperator. The LME~\eqref{eq:LME_sys} is written in a physically intuitive form facilitating later use, with $J\rho J^\dag$ representing emission events (quantum jumps) into the waveguide and ${\cal L}_0\rho$ the evolution conditioned on the absence of such jumps. 

Continuous sensing relies on measuring the output photons in the waveguide, the quantum state thereof via tracing away the sensor and all environments from the global state, $\Xi(\theta,T):={\rm tr}_{\rm S,E}[\ket{\Psi(\theta,T)}\bra{\Psi(\theta,T)}]$, leading formally to a cMPO
\begin{align}
\label{eq:rhoE}
\Xi(\theta,T)={\rm tr}_{\rm S}\left\{{\cal T}e^{ \int_0^Tdt[{\cal L}_0(\theta)+{\cal J}(\theta){\cal G}(t)+{\cal K}(\theta){\cal F}(t)]}\rho(0)\right\}\Xi(0).
\end{align}
Here ${\cal J,K}$ are (potentially $\theta$-dependent) superoperators of the sensor, ${\cal J}(\rho):=J\rho$, ${\cal K}(\rho):=\rho J^\dag$; ${\cal G}(t), {\cal F}(t)$ are superoperators of the waveguide, ${\cal G}(t)(\Xi):=f^\dag(t)\Xi$, ${\cal F}(t)(\Xi):=\Xi f(t)$. 
{While Eq.~\eqref{eq:rhoE} was derived assuming a specific initial global state $\Ket{\Psi(0)}=\ket{\psi_{\rm S}(0)}\otimes\ket{\rm vac}_{\rm W,E}$, it remains valid for an arbitrary sensor initial state $\rho(0)$, provided that the waveguide input is a vacuum (or coherent)-state,  $\Xi(0)=\ket{\rm vac}_{\rm W}\bra{\rm vac}$, and the environments are Markovian and Gaussian. Typical environments include vacuum environments, where ${\cal L}_0$ follows Eq.~\eqref{eq:no_jump_evol}, and thermal environments, where ${\cal L}_0$ is modified by replacing ${\cal D}[L_k]$ with $(n_k+1) {\cal D}[L_k]+n_k{\cal D}[L_k^\dag]$ in Eq.~\eqref{eq:no_jump_evol}, with $n_k$ the thermal occupation.}

\emph{Quantum Fisher Information \& Bargmann invariants of the Waveguide.---}Below we develop an efficient method to evaluate the QFI, $I(\theta,T)$, of the waveguide density operator Eq.~\eqref{eq:rhoE}. For unbiased estimation of $\theta$, the QFI quantifies the minimal MSE~\footnote{The MSE reduces to the variance if the estimator is unbiased.}, ${\rm min}[\langle(\theta_{\rm est}-\theta)^2\rangle]=1/[N I(\theta,T)]$, optimized over $N\gg1$ repetitions of arbitrary (including adaptive) measurement of the sensor output photons for the time window $[0,T]$. 
The QFI can be formally expressed as $I(\theta,T)=\sum_{ \lambda+\lambda^\prime>0}\frac{2}{{\lambda}+{\lambda^\prime}}\left| \langle \lambda \left|\partial _{\theta}\Xi(\theta,T)\right|\lambda^\prime\rangle\right|^2$~\cite{Helstrom1976,PhysRevLett.72.3439}, where $\{\lambda,\ket{\lambda}\}$ forms the eigensystem of the waveguide density operator Eq.~\eqref{eq:rhoE}, $\Xi(\theta,T)=\sum_\lambda \lambda\ket{\lambda}\bra{\lambda}$. Such eigendecomposition, however, is infeasible to compute due to the continuous, infinite-dimensional nature of Eq.~\eqref{eq:rhoE}. 

Instead, our method is based on relating the QFI to nonlinear powers of the waveguide density operator known as the Bargmann invariants~\cite{10.1063/1.1704188},
\begin{equation}
\label{eq:Bargmann}
B({\Theta},T):={\rm tr}[\Xi(\theta_1,T)\Xi(\theta_2,T)\dots\Xi(\theta_{m+2},T)],
\end{equation}
with ${\Theta}:=\{\theta_1,\theta_2,\dots,\theta_{m+2}\}$ and $m\geq0$. We show in \cite{supplement} that $I(\theta,T)$ is the limit of a series $\{I_n(\theta,T)\mid n\in\mathbb{N}_0\}$, which increases monotonically and converges exponentially as $I(\theta,T)-I_n(\theta,T)={\cal O}(\xi^{n})$, with $\xi\equiv1-{\rm min}[\lambda]/\sqrt{\sum_\lambda \lambda^2}$. The elements of the series can be conveniently expressed in terms of the Bargmann invariants Eq.~\eqref{eq:Bargmann} as
\begin{subequations}
\begin{align}
I_{n}(\theta,T)=&\sum_{m=0}^n (-1)^m\binom{n+1}{m+1}\frac{f_m(\theta,T)}{\Lambda^{m+1}(\theta,T)}\label{eq:In}\\
f_m(\theta,T)=&\sum_{l=0}^m {D^{m}_l}\partial_{\mu,\nu}^2{\rm tr}[\Xi(\mu,T)^{l+1}\Xi(\nu,T)^{m-l+1}]|_{\mu=\nu=\theta},\label{eq:fm}
\end{align}
\end{subequations}
where $\Lambda(\theta,T)=2\{{\rm tr}[\Xi^2(\theta,T)]\}^{1/2}$, and $D^{m}_l=2\binom{m}{l}-\binom{m}{l+1}-\binom{m}{l-1}$ with the convention $\binom{m}{l}=0$ for $l<0$ or $l>m$. Remarkably, although being nonlinear functionals of the waveguide density operator, the Bargmann invariants Eq.~\eqref{eq:Bargmann} can be efficiently evaluated via extension of theoretical techniques of open quantum systems, as we establish below. Efficient evaluation of Eq.~\eqref{eq:Bargmann} hence unlocks, via finite difference approximation of the derivatives in Eqs.~\eqref{eq:In} and \eqref{eq:fm}, the access to the QFI.

\emph{Generalized Replica Master Equation.---}{As a key result of the present Letter, the waveguide Bargmann invariants Eq.~\eqref{eq:Bargmann} can be evaluated systematically from the \emph{reduced dynamics} of the sensor alone, via}
\begin{equation*}
{B({\Theta},T)=\text{tr}\varrho(\Theta,T),}
\end{equation*}
where $\varrho(\Theta,T)$ is an operator acting on $m+2$ replicas of the sensor. It satisfies the initial condition $\varrho(\Theta,0)=\otimes_{\alpha=1}^{m+2}\rho_{\rm}^{(\alpha)}(0)$, with $\rho(0)$ the sensor initial density operator, and evolves according to~\cite{supplement}
\begin{align}
\label{eq:replicaME}
\dot{\varrho}(\Theta,t)=&\sum_{\alpha =1}^{m+2}\mathcal{L}_{0}^{(\alpha )}(\theta_\alpha)\varrho(\Theta,t)
\nonumber\\
&+\sum _{\alpha =1}^{m+2}J^{(\alpha+1)}(\theta _{\alpha+1})\varrho(\Theta,t)J^{(\alpha)\dagger}(\theta_{\alpha}),
\end{align}
with operators with superscript $(\alpha)$ acting on $\rho^{(\alpha)}$, and periodic boundary condition imposed on the replica index, $(m+3)\equiv (1)$. Equation \eqref{eq:replicaME} is not a LME and does not describe a completely positive trace-preserving (CPTP) map; hereafter we call it a $(m+2)$-GRME. Its first line describes independent evolution for each replica according to Eq.~\eqref{eq:no_jump_evol}, i.e., in the absence of emission into the waveguide. The second line of Eq.~\eqref{eq:replicaME} describes collective emission events (quantum jumps) into the waveguide between adjacent replicas, which build up inter-replica correlations. Consequently, $\varrho(\Theta,t)$ becomes non-factorizable with respect to the replicas for $t>0$. 

Direct numerical propagation of Eq.~\eqref{eq:replicaME}, although implementable for a small replica number, quickly becomes infeasible with increasing $m$---the storage cost being ${\cal O}(D^{2m+4})$ (with $D$ the sensor Hilbert space dimension), and the propagation cost (i.e., the number of multiplications) ${\cal O}(TD^{4m+8})$. Instead, we develop a numerically efficient, scalable approach to integrate Eq.~\eqref{eq:replicaME} based on the time-evolving block decimation (TEBD) algorithm. We Trotterize the infinitesimal-time propagator of Eq.~\eqref{eq:replicaME} into $m+2$ gates acting on adjacent replicas $\alpha$ and $\alpha+1$, 
\begin{align}
\label{eq:TEBD}
\varrho(\Theta,t+dt)=&\prod_{\alpha=1}^{m+2}V_{\alpha,\alpha+1}(dt) \varrho(\Theta,t)+{\cal O}(dt^2), \nonumber\\
V_{\alpha,\alpha+1}(dt)\varrho=&\exp\bigg\{J^{(\alpha+1)}(\theta_{\alpha+1})\varrho J^{(\alpha)\dag}(\theta_{\alpha})dt\nonumber\\
&+\frac{dt}{2}\left[{\cal L}_0^{(\alpha)}(\theta_{\alpha})+{\cal L}_0^{(\alpha+1)}(\theta_{\alpha+1})\right]\varrho\bigg\}.
\end{align}
These gates can be applied efficiently to the replica chain based on matrix-product representation of $\varrho(\Theta,t)$ across the chain~\cite{supplement}, leading to a significant reduction of the storage cost to ${\cal O}[(m+2)\chi^2D^2]$, and the computational cost to ${\cal O}[(m+2)T\chi^3 D^{6}]$, with $\chi$ the bond dimension of the matrix-product representation. Efficient propagation is achieved whenever $\chi$ can be kept small.

Crucially, the inter-replica correlation in Eqs.~\eqref{eq:replicaME} and \eqref{eq:TEBD} is established solely via the collective quantum jumps $\propto J^{(\alpha+1)}\varrho J^{(\alpha)\dag}$ which, remarkably, also suppress entanglement growth by quenching the excitation of the replicas involved. As detailed in~\cite{supplement}, we observe that this dissipative nature leads to an \emph{area law} of the bipartite operator entanglement entropy across the replica chain for all propagation time $T$, in sharp contrast to the volume law typically established by unitary entangling gates
~\cite{oliveira2007generic,dahlsten2007emergence}. Consequently, TEBD can be performed for the GRME \eqref{eq:replicaME} with a small $\chi$ over sufficiently large $m$ and $T$, enabling the efficiency and scalability of our method~\cite{VerstraeteC06}. 

\begin{figure}[b!]
\centering{} \includegraphics[width=0.48\textwidth]{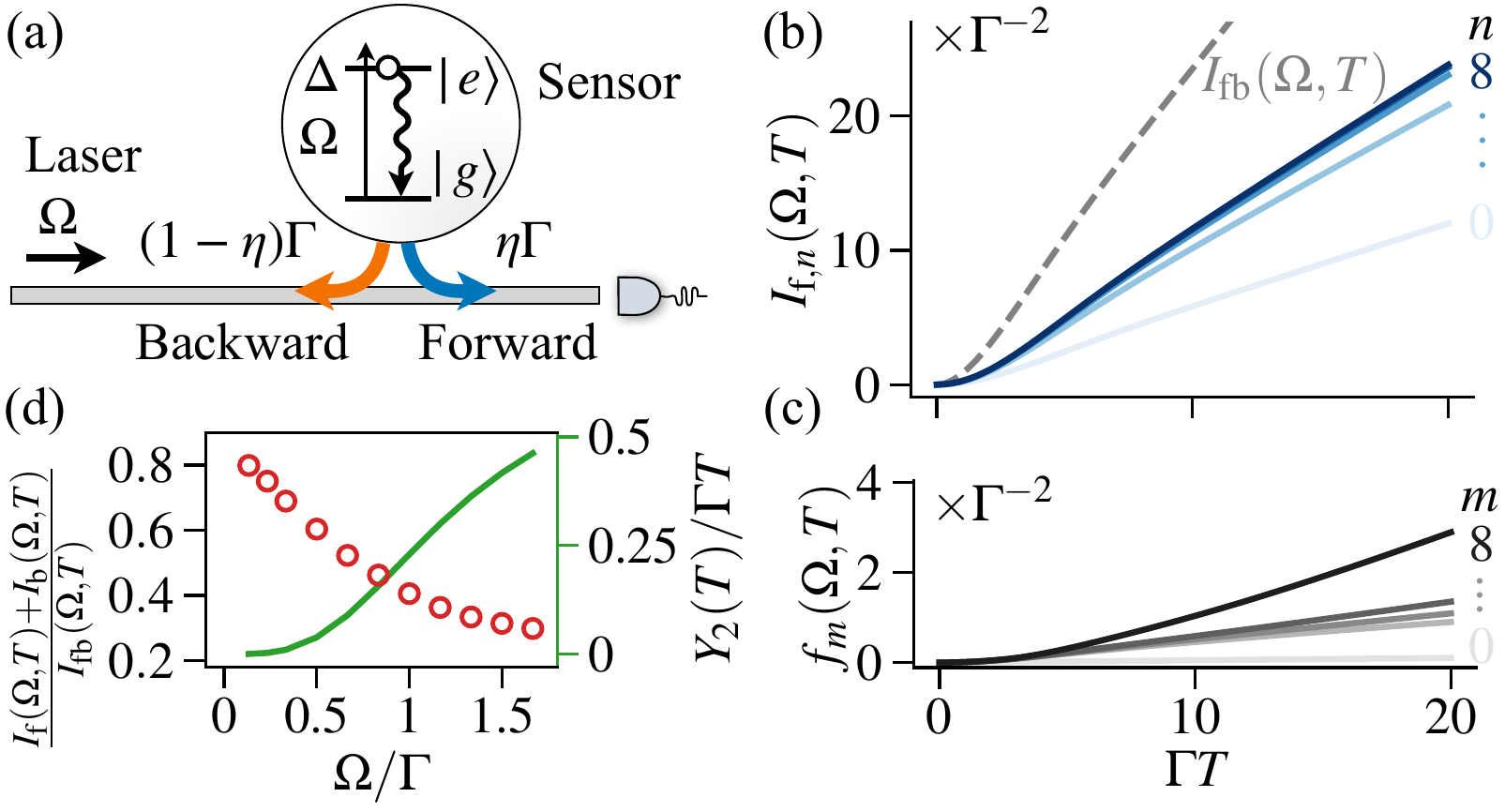} 
\caption{Illustration of the method. (a) The sensor is an emitter driven at Rabi-frequency $\Omega$ and detuning $\Delta$, and interfaced bidirectionally with a waveguide with a forward/backward emission rate $\eta\Gamma/(1-\eta)\Gamma$. (b) The approximate QFI of forward-emitted photons for Rabi-frequency estimation, $I_{{\rm f},n}(\Omega,T)$, for $n=0,2,4,6,8$ (light to dark), showing excellent convergence at $n=8$. Dashed curve shows $I_{\rm fb}(\Omega,T)$, the joint QFI of forward and backward emitted photons, evaluated using the method of~\cite{PhysRevX.13.031012}. Parameters: $\Delta=\Omega=\Gamma/3$, $\eta=2/3$. (c) The Bargmann derivatives [see Eq.~\eqref{eq:fm}] for forward-emitted photons at $m=0,2,4,6,8$ (light to dark), evaluated at the same parameters as (b). (d) The forward and backward emitted photons become increasingly entangled as $\Omega$ increases, as reflected by the unit-time Renyi mutual information $Y_2(T)/T$. Consequently, the QFI retrievable via independent measurements of forward and backward emitted photons, $I_{\rm f}(\Omega,T)+I_{\rm b}(\Omega,T)$, becomes increasingly small compared to the joint QFI $I_{\rm fb}(\Omega,T)$. Parameters: $\Delta=\Gamma/3,\eta=2/3$.
}
\label{fig:fig2} 
\end{figure}

\emph{Illustration.---}As a first example, we apply our method to a typical lossy quantum optical setup: a driven sensor coupled to two photonic channels, f and b, with photon detection limited to channel f. This models, for instance, a nonchiral emitter-waveguide interface, where f(b) represents forward (backward) propagating modes, or atom-laser scattering, with f(b) corresponding to laser (free-space) modes. Consider a driven two-level sensor (TLS) with excited (ground) state $\ket{e(g)}$, see Fig.~\ref{fig:fig2}(a), whose Hamiltonian in the rotating frame reads $H=-\Delta \sigma_{ee}+(\Omega \sigma_{eg}+{\rm h.c.})/2$, with $\Delta$($\Omega$) the laser detuning (Rabi frequency) and $\sigma_{ij}:=\ket{i}\bra{j}$. The sensor couples to the channel f and b via the jump operator $J_{\rm f}=\sqrt{\eta\Gamma}\sigma_{ge}$ and $J_{\rm b}=\sqrt{(1-\eta)\Gamma}\sigma_{ge}$ respectively, with $\eta\leq1$ the photon collection efficiency and $\Gamma$ the total emission rate.

Consider sensing the Rabi frequency $\Omega$ through forward photon detection. Figure~\ref{fig:fig2}(c) shows the corresponding $f_m(\Omega,T)$ computed from Eq.~\eqref{eq:fm} via integrating the GRME~\eqref{eq:replicaME}, assuming $\rho(0)=\ket{g}\bra{g}$. From this, the approximate QFI $I_{{\rm f},n}(\Omega,T)$ is obtained via Eq.~\eqref{eq:In}, which exhibit rapid convergence with increasing $n$, achieving accurate estimate of $I_{\rm f}(\Omega,T)$ at $n\simeq8$, cf. Fig.~\ref{fig:fig2}(b). The QFI displays an initial transient behavior for $T\lesssim 1/\Gamma$; and transitions to linear growth with $T$ for $T\gg 1/\Gamma$, i.e., as the sensor steady state is reached. 

The QFI of the backward emitted photons, $I_{\rm b}(\Omega,T)$, can be computed similarly. The sum of $I_{\rm f}(\Omega,T)$ and $I_{\rm b}(\Omega,T)$ represents the optimal precision obtainable from \emph{independent} measurements of both channels. We can however regard f and b as a single channel and compute the corresponding QFI $I_{\rm fb}(\Omega,T)$, which quantifies the optimal precision obtainable from any \emph{joint} measurement of both. In Fig.~\ref{fig:fig2}(d) we show their ratio, $[I_{\rm f}(\Omega,T)+I_{\rm b}(\Omega,T)]/I_{\rm fb}(\Omega,T)$, for $T\gg 1/\Gamma$, which monotonously decreases with increasing $\Omega$, indicating growing entanglement between the two fields. Such entanglement can be quantified by the inter-field Renyi mutual information, $Y_{\ell}(T):=S_{\ell,{\rm f}}(T)+S_{\ell,{\rm b}}(T)-S_{\ell,{\rm fb}}(T)$, with $S_{\ell,i}(T)={\rm log}_{\ell}[{\rm tr}(\Xi_i^\ell)]/(1-\ell)$ the $\ell$-th order Renyi entropy of field $i\in\{{\rm f, b}\}$,  that is readily computable via the Bargmann invariants Eq.~\eqref{eq:Bargmann}. For $T\gg 1/\Gamma$, $Y_{\ell=2}(T)/T$ is shown in Fig.~\ref{fig:fig2}(d).

\begin{figure}[t!]
\centering{} \includegraphics[width=0.48\textwidth]{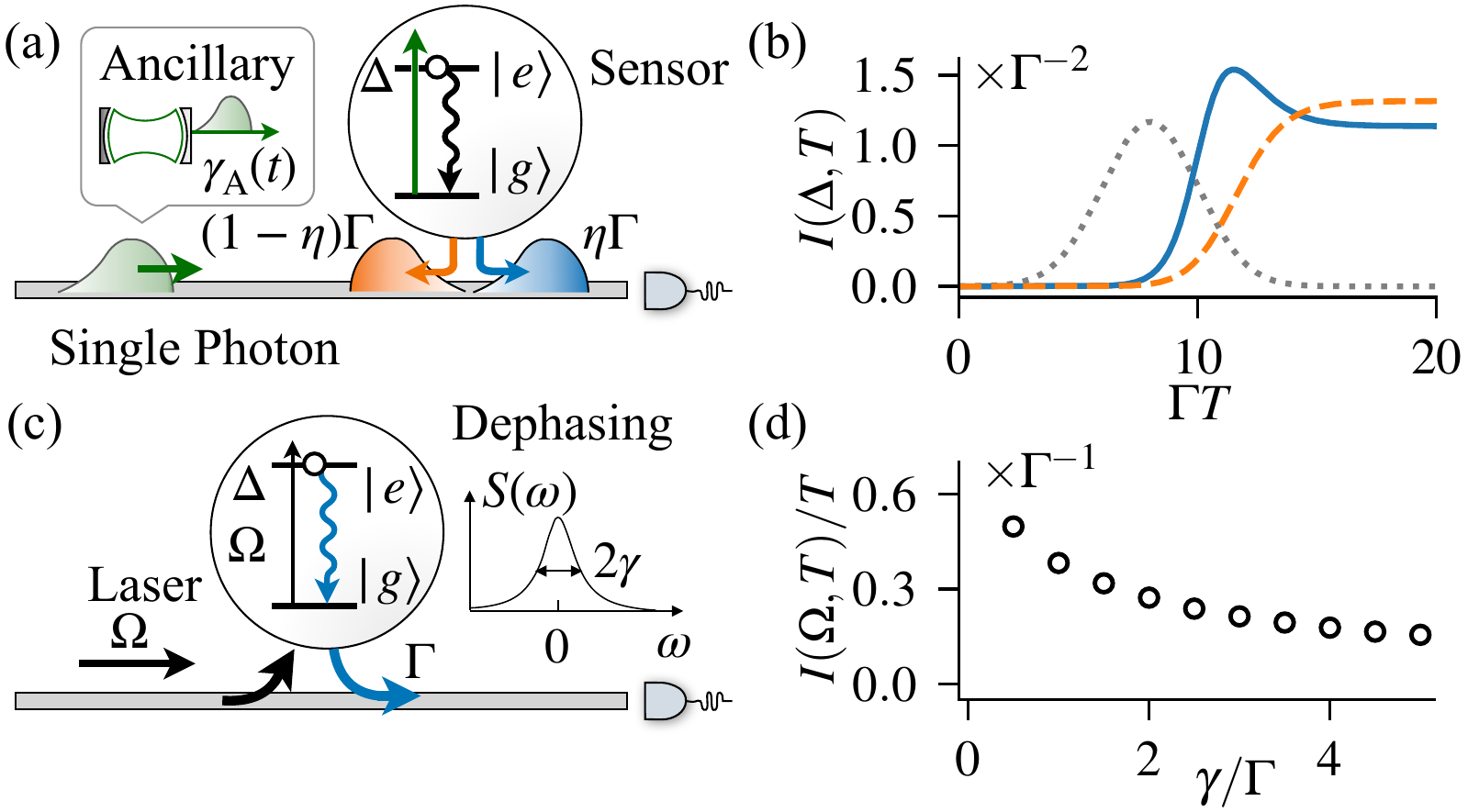} 
\caption{Applications of the extended method. (a) A bidirectional sensor-waveguide interface with a continuous-mode single-photon input. The input field can be generated via an ancillary cavity mode (see texts). (b) The temporal shape of the input photon (doted, grey) $u(t)={\rm exp}[-(t-t_0)^2/2\tau^2]/\sqrt{\tau}\pi^{1/4}$, and the QFI of the forward (solid, blue) and backward (dashed, orange) propagating fields. Parameters: $t_0=8/\Gamma$, $\tau=1/\Gamma,\Delta=\Gamma/4,\eta=1/2$, the sensor is initialized in $\ket{g}$. (c) A unidirectional sensor-waveguide interface, with the sensor subjected to non-Markovian dephasing. The environment has a Lorentzian noise spectrum $S(\omega)=\gamma^2/(\omega^2+\gamma^2)$ with the spectral width $\gamma$ adjustable. (d) The QFI of the emission field as a function of $\gamma$. Parameters: $\Delta=0,\Omega=\Gamma$, the sensor is initialized in $\ket{g}$.
}
\label{fig:fig3} 
\end{figure}

\emph{Extension.---}Our method naturally applies to sensors under time-varying driving, by allowing ${\cal L}_0(\theta,t),J(\theta,t)$ in Eq.~\eqref{eq:replicaME} to vary with time. Furthermore, although the GRME~\eqref{eq:replicaME} is formulated for sensors driven by a coherent input field, our method is readily extendable to accommodate arbitrary input quantum field---by introducing an ancillary open system upstream of the sensor, see~Fig.~\ref{fig:fig3}(a), and designing the input field of the sensor through controlling the dynamical parameters, i.e., the Hamiltonian $H_{\rm A}(t)$ and jump operator $J_{\rm A}(t)$, of the ancilla~\cite{PhysRevA.107.013705}. Correspondingly, the QFI of the sensor output field and the associated Bargmann invariants can be evaluated with Eq.~\eqref{eq:replicaME} for the cascaded ancilla-sensor setup, as constructed via replacements ${\cal L}_0(\theta){\varrho}\to{\cal L}_0(\theta){\varrho}-i[H_{\rm A}(t)+H_{\rm casc}(t),\varrho]$ and $J(\theta)\to J(\theta)+J_{\rm A}(t)$, with $H_{\rm casc}(t)=(i/2)[J(\theta)J_{\rm A}^\dag(t)-J_{\rm A}(t) J^\dag(\theta)]$. 

As an illustration, we consider the bidirectional TLS-waveguide interface as before, with the waveguide input consisting of a continuous-mode single-photon wave-packet $\ket{\mathbf{1}_u}=\int dt u(t) f^\dag(t)\ket{\rm vac}_{\rm W}$, as relevant to lossy quantum light spectroscopy experiments~\cite{Mukamel_2020,PhysRevA.107.062601}. The input photon can be generated with an upstream ancillary cavity mode $(a,a^\dag)$ initialized in the first Fock state, $\rho_{\rm A}(0)=\ket{1}_{\rm A}\bra{1}$, with $H_{\rm A}(t)=0$ and $J_{\rm A}(t)=\sqrt{\gamma_{\rm A}(t)}a$, wherein $\gamma_{\rm A}(t)$ is related to the photonic temporal profile as $\gamma_{\rm A}(t)= u^2(t)/[1-\int_0^t d\tau u^2(\tau)]$~\cite{PhysRevA.107.013705}. Evolving the cascaded GRME provides us with the QFI of both forward and backward propagating output-fields, as shown in Fig.~\ref{fig:fig3}(b). 

Our framework further extends to continuous sensors subject to non-Markovian noise by making use of the pseudomode representation~\cite{PhysRevLett.120.030402,doi:10.1142/S1230161222500196,PhysRevLett.132.100403} of structured environments. The representation replaces the original structured environment by a small number of damped bosonic modes interacting with the sensor, whose parameters are chosen to reproduce the expectation values and correlation functions of the original environment. We show in~\cite{supplement} that the QCRB under non-Markovian noise can be evaluated by propagating the GRME~\eqref{eq:replicaME} for the composite system consisting of the open sensor and the pseudomodes. As an illustrative application of this theoretical extension, Fig.~\ref{fig:fig3}(c-d) presents the QFI of the output field of a unidirectional TLS-waveguide interface, with the sensor dephased by an environment with a Lorentzian noise spectrum~\cite{supplement}. Finally, although we have focused on estimating a single constant parameter, our framework readily extends to waveform estimation; see End Matter.

\emph{Conclusions.}---We have established a framework for efficient evaluation of the quantum Cram\'er-Rao bound (QCRB) of continuously monitored quantum sensors subjected to general noise. Our framework extends further to computing general properties of quantum-optical input-output channels that are nonlinear in the field density operator~\cite{Dayou}. 
Future directions include extension to noisy hypothesis testing, and designing optimal measurement to saturate the QCRB. A promising approach for the latter is the recently established universal quantum noise cancellation strategy~\cite{PhysRevX.13.031012,Godley2023adaptivemeasurement,Tsang2025quantumreversal}, for which our present work provides a rigorous benchmark for assessing the optimality under realistic noise and imperfections. 

\emph{Acknowledgements.}---We thank {Klaus M{\o}lmer for helpful comments
on the manuscript, and} Francesco Albarelli, Animesh Datta, Aiman Khan and Gabriela W\'ojtowicz for discussions. This work was supported by the ERC Synergy grant HyperQ (Grant No.~856432), the EU projects QuMicro (Grant No.~101046911) and C-QuENS (Grant No.~101135359), the BMBF project PhoQuant (Grant No.~13N16110), and the start-up funds for D. Y. from HFNL. We acknowledge support by the state of Baden-Württemberg through bwHPC and the German Research Foundation (DFG) through Grant No. INST 40/575-1 FUGG (JUSTUS 2 cluster). Part of the numerical simulations were performed using the QuTiP library~\cite{JOHANSSON20131234}.

\emph{Data Availability.}---The data that support the findings of this article are openly available~\cite{yang_2026_18365501}. Example codes are available from the authors upon reasonable request.

\emph{Note Added.}---While completing the manuscript, we became aware of a related work~\cite{ljh3-3l4j} on variational calculation of the precision limit of lossy quantum light spectroscopy. Both manuscripts have been synchronized to appear on arXiv on the same date.
\\\\
\noindent\textbf{{End Matter}}

\emph{Lossy QCRB: linear vs. nonlinear sensors.}---Consider the scenario in Fig.~\ref{fig:fig2}(a), where a sensor is driven by a coherent input, and emits photons into the accessible channel (labeled as f) and the inaccessible channel (labeled as b) at a rate $\eta\Gamma$ and $(1-\eta)\Gamma$ respectively. An intriguing and practically relevant question is, how does the QFI of the accessible photons, $I_{\rm f}(\theta,T)$, depend on the efficiency $\eta$? As a crude guess, one might estimate $I_{\rm f}(\theta,T)$ as $\eta I_{\rm fb}(\theta,T)$, where $I_{\rm fb}(\theta,T)$ is the joint QFI of the accessible and inaccessible photons. Leveraging our numerical framework, here we elucidate this question by quantitative computation of model systems, providing useful insights for real-world sensor platforms.

Specifically, focusing on the estimation of the driving strength $\Omega$ (see definition below), we obtain the following results: (i) $\eta I_{\rm fb}(\theta,T)$ is indeed a tight bound, i.e., $I_{\rm f}(\theta,T)=\eta I_{\rm fb}(\theta,T)$, for lossy \emph{linear} amplifiers (as excellent models, e.g., of optomechanical force sensors~\cite{Barzanjeh:2022ve} and ensemble magnetometers~\cite{Moller:2017ta}); (ii) For nonlinear continuous sensors, such as the lossy TLS studied in the main text (for which $\Omega$ becomes the Rabi frequency), $\eta I_{\rm fb}(\theta,T)$ is not an accurate measure of the retrievable QFI. Interestingly enough, for the lossy TLS we observe that it serves as a loose \emph{upper bound} to the QFI, $I_{\rm f}(\theta,T)\leq\eta I_{\rm fb}(\theta,T)$.

A linear amplifier is a continuous variable system with a pair of conjugate quadratures $X=(a^\dag+a)/\sqrt{2}$ and $P=i(a^\dag-a)/\sqrt{2}$ satisfying $[a,a^\dag]=1$. These can, e.g., be the position and momentum operators of a mechanical oscillator, or be the collective internal spins along two orthogonal directions of an atomic ensemble~\cite{Moller:2017ta}. The unknown force $\Omega$ couples with $X$, resulting in the sensor Hamiltonian
\begin{equation}
\label{eq:linearamplifierH}
H=\omega a^\dag a - \Omega(a+a^\dag),
\end{equation}
with $\omega$ the oscillation frequency. 

The sensor further couples to the detectable and undetectable photonic channels via the jump operator $\sqrt{\eta\Gamma_{\rm}}X$ and $\sqrt{(1-\eta)\Gamma_{\rm}}X$ respectively. For an optomechanical force sensor~\cite{Barzanjeh:2022ve}, such a coupling is mediated by a driven damped cavity mode $(c, c^\dag)$, which interacts with the oscillator via the linearized optomechanical coupling $V= g (c^\dag + c) X$. The cavity damping rate is typically much larger than $g$, allowing us to adiabatically eliminate the cavity mode hence to arrive at the aforementioned quadrature jump operators. For an atomic gas magnetometer, such coupling is enabled by the Faraday rotation~\cite{Moller:2017ta,Colangelo:2017wi}. 

\begin{figure}[t!]
\centering{} \includegraphics[width=0.48\textwidth]{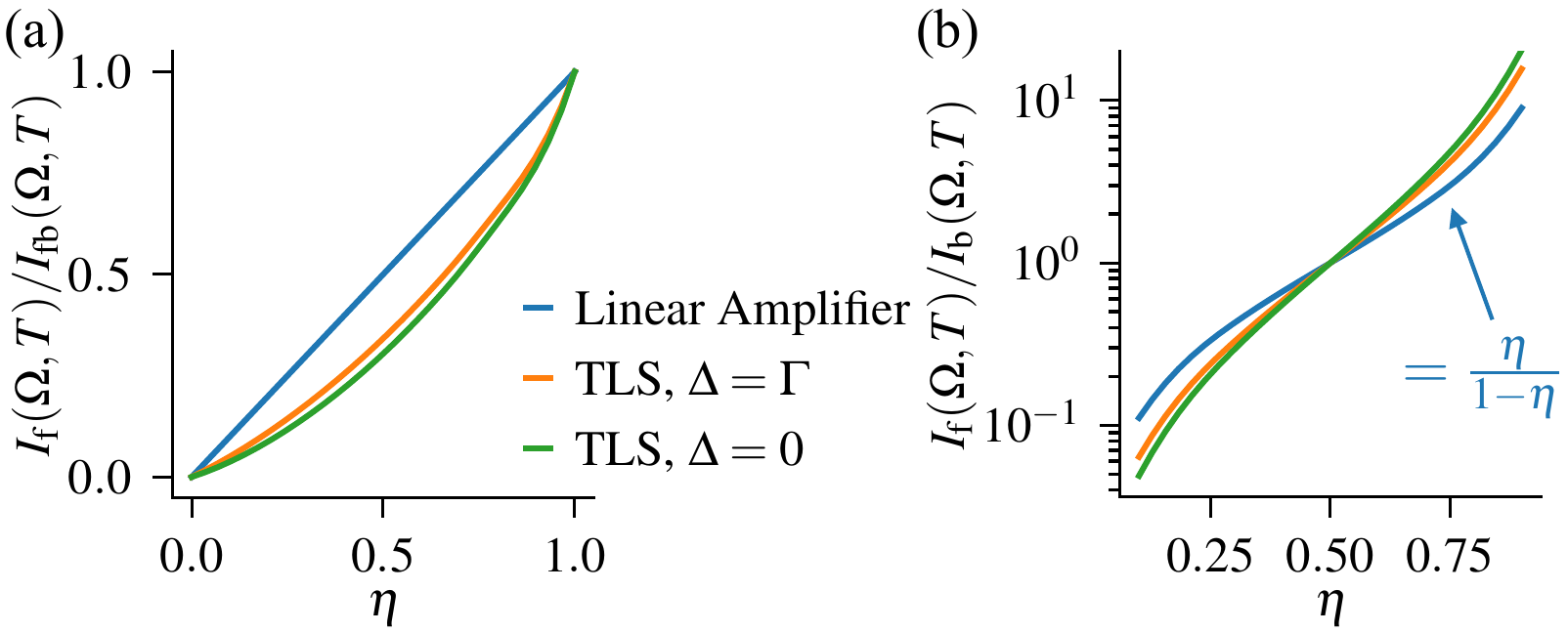} 
\caption{{Comparison between the quantum Cram\'er–Rao bound (QCRB) for a linear amplifier and a two-level sensor (TLS), assuming finite photon collection efficiency $\eta<1$. (a) The QFI carried by the accessible photons, $I_{\rm f}(\Omega,T)$, normalized by the total QFI of the emitted photons $I_{\rm fb}(\Omega,T)$, for estimating the parameter $\Omega$ with each sensors. In the numerical calculation we choose for the amplifier $\omega=\Omega=\Gamma/3$, and for the TLS $\Omega=\Gamma/3$ and $\Delta$ adjustable. (b) The ratio between the retrievable and lost information, $I_{\rm f}(\Omega,T)/I_{\rm b}(\Omega,T)$, as a function of $\eta$, for both the linear amplifier and the TLS. Parameters are the same as panel (a).
}}
\label{fig:fig4} 
\end{figure}
Using our numerical method, we evaluate the retrievable information for sensing the driving strength $\Omega$, $I_{\rm f}(\Omega,T)$, for various detection efficiency $\eta$, for both the lossy linear amplifier [cf. Eq.~\eqref{eq:linearamplifierH}] and the lossy TLS (cf. the main text). As demonstrated in Fig.~\ref{fig:fig4}(a), $I_{\rm f}(\Omega,T)$ precisely matches $\eta I_{\rm fb}(\Omega,T)$ for lossy linear amplifiers, confirming its tightness; in contrast, we observe that $\eta I_{\rm fb}(\Omega,T)$ serves only as a loose upper bound for the QFI of the lossy TLS---it becomes tight only for $\eta=0$ or $1$. While Fig.~\ref{fig:fig4} is shown for specific parameters, we have checked numerically that the observed behavior persists for other parameters as well.

We can similarly verify, for linear amplifiers, that the QFI carried by the inaccessibe photons is $I_{\rm b}(\Omega,T)=(1-\eta) I_{\rm fb}(\Omega,T)$. In Fig.~\ref{fig:fig4}(b), we show the ratio between the retrievable and lost information, $I_{\rm f}(\Omega,T)/I_{\rm b}(\Omega,T)$, for both a linear amplifier and a TLS. The linear-amplifier ratio agrees exactly with $\eta/(1-\eta)$; the TLS does not.

Our numerics lead us to conjecture that $\eta I_{\rm fb}(\theta,T)$ is not an accurate quantifier of the precision for sensing the coherent driving strength with generic nonlinear continuous sensors~\cite{PRXQuantum.3.010354,Ding:2022aa,doi:10.1126/sciadv.ado8130,Ilias:2024aa,PhysRevLett.132.050801}, where the accessible and inaccessible photons develop intricate non-Gaussian entanglement. Whether it provides a loose upper bound for the QFI, as in the TLS case, is an interesting open question for future study.

\emph{Extension to Waveform Estimation.}---Our framework extends naturally to evaluating the QCRB for waveform estimation in the presence of noise. Therein, an open quantum sensor is subject to a time-varying waveform $\vartheta(t),t\in[0,T]$, and monitored continuously. An estimator waveform $\vartheta_{\rm est}(t)$ is then constructed from the continuously recorded signals $D(0,T)$---a task involving joint estimation of infinitely many parameters. To make the estimation problem well defined, a prior distribution functional $P[\vartheta]$ (throughout this section, bracketed expressions of the form $f[\cdot]$ denote functionals) reflecting known constraints on the waveform distribution (e.g., its finite bandwidth) is essential~\cite{PhysRevLett.106.090401}.

Consider unbiased estimation of $\vartheta(t)$. The QCRB for this case is expressed in terms of the matrix inequality 
\begin{equation}
\label{eq:QCRBwaveform}
\Sigma(t,t')-{\cal I}^{-1}(t,t')\geq 0,
\end{equation}
where $\Sigma(t,t')$ is the weighted error matrix~\cite{PhysRevLett.106.090401},
\begin{align}
\Sigma(t,t'):=&\int \mathfrak{D}[\vartheta,D(0,T)]P[\vartheta]P[D(0,T)|\vartheta]\nonumber\\
&\times\Big(\vartheta_{\rm est}(t)-\vartheta(t)\Big)\Big(\vartheta_{\rm est}(t')-\vartheta(t')\Big)\nonumber
\end{align}
with $\int \mathfrak{D}[\vartheta,D(0,T)]$ a (functional) integral over all possible waveforms $\vartheta(t)$ and records $D(0,T)$, ${\cal I}$ the QFI matrix and ${\cal I}^{-1}$ its inverse [defined via $\int_0^T ds {\cal I}(t,s){\cal I}^{-1}(s,t')=\delta(t,t')$]. In particular, Eq.~\eqref{eq:QCRBwaveform} implies bounds to the point MSE for unbiased estimators,
\begin{equation}
\langle[\vartheta_{\rm est}(t)-\vartheta(t)]^2\rangle\geq {\cal I}^{-1}(t,t).
\end{equation} 

Ref.~\cite{PhysRevLett.106.090401} shows that $\cal I$ is given by a sum of quantum and classical contributions, $\cal I=I^{(Q)}+I^{(C)}$, with
\begin{align}
{\cal I^{(C)}}(t,t')=\int \mathfrak{D}[\vartheta] P[\vartheta]\frac{\delta P[\vartheta]}{\delta \vartheta(t)}\frac{\delta P[\vartheta]}{\delta \vartheta(t')}
\end{align}
the Fisher information matrix of the prior distribution $P[\vartheta]$, and $\delta/\delta\vartheta(t)$ a functional derivative. The other term is the weighted QFI matrix for the waveguide photons,
\begin{subequations}
\begin{align}
{\cal I^{(Q)}}(t,t')&=\int \mathfrak{D}[\vartheta]P[\vartheta]\tilde{\cal I}(t,t'),\label{eq:QFI_waveform}\\
\tilde{\cal I}(t,t')&=\frac{1}{2}\sum_{\lambda,\lambda'}(\lambda+\lambda'){\rm Re}[\langle \lambda| L_t |\lambda^\prime\rangle\langle\lambda^ \prime |L_{t'}|\lambda\rangle],\label{eq:QFImat}
\end{align}
\end{subequations}
where Re denotes the real part, $\{\lambda,\ket{\lambda}\}$ forms the eigensystem of the waveguide density operator $\Xi[\vartheta(t)]$, and the symmetric logarithmic derivative operator $L_t$ is defined as
\begin{align}
\frac{\delta\Xi[\vartheta(t)]}{\delta\vartheta(t)}=\frac{1}{2}(L_t\Xi+\Xi L_t).
\end{align}
Evaluation of the QCRB Eq.~\eqref{eq:QCRBwaveform} hence reduces to the efficient calculation of $\tilde{{\cal I}}(t,t')$, based on the knowledge of the sensor dynamics and the noise model. Remarkably, this can be achieved via direct extension of our method.

Specifically, as we derive in~\cite{supplement}, $\tilde{\cal I}(t,t')$ can be evaluated as the limit of an exponentially converging matrix series
$\{\tilde{\cal I}_n\mid n\in\mathbb{N}_0\}$, with
\begin{subequations}
\begin{align}
\tilde{{\cal I}}_n(t,t')=&\sum_{m=0}^n (-1)^m\binom{n+1}{m+1}\frac{f_m(t,t')}{\Lambda^{m+1}[{\vartheta}]},\label{eq:InWeveform}\\
f_m(t,t')=&\sum_{l=0}^m {D^{m}_l}\frac{\delta^2}{\delta\mu(t)\delta\nu(t')}{\rm tr}\!\left(\Xi[\mu]^{l+1}\Xi[{\nu}]^{m-l+1}\right)\!|_{{\mu}={\nu}={\vartheta}},\label{eq:fmWeveform}
\end{align}
\end{subequations}
Here, $\Lambda[\vartheta]:=2\{{\rm tr}\Xi^2[\vartheta]\}^{1/2}$ and $D_l^m$ is defined in the main text [below Eq.~\eqref{eq:fm}]. The functional derivatives in Eq.~\eqref{eq:fmWeveform} can be implemented numerically and robustly via the finite element method~\cite{Reddy19}. All the elements in Eqs.~\eqref{eq:InWeveform} and \eqref{eq:fmWeveform} are hence accessible via the waveform-parameterized Bargmann invariants,
\begin{equation}
\label{eq:BargmannWaveform}
B[{\tilde{\Theta}}]:={\rm tr}\{\Xi[\vartheta_1]\Xi[\vartheta_2]\dots\Xi[\vartheta_{m+2}]\}\equiv {\rm tr}\{\varrho[\tilde{\Theta}]\},
\end{equation}
with $\tilde{\Theta}:=\{\vartheta_1,\vartheta_2,\dots,\vartheta_{m+2}\}$ being a set of $m+2$ waveforms, and $\varrho[\tilde{\Theta}]$ evolving according to the waveform-parameterized GRME
\begin{align}
\label{eq:replicaMEwaveform}
\dot{\varrho}[\tilde{\Theta}]=&\sum_{\alpha =1}^{m+2}\mathcal{L}_{0}^{(\alpha)}[\vartheta_\alpha]\varrho[\tilde{\Theta}]
\nonumber\\
&+\sum _{\alpha =1}^{m+2}J^{(\alpha+1)}[\vartheta _{\alpha+1}]\varrho[\tilde{\Theta}]J^{(\alpha)\dagger}[\vartheta_{\alpha}].
\end{align}

The above extension provides a general framework for evaluating the noisy precision limit for waveform estimation~\cite{PhysRevLett.106.090401,PhysRevLett.111.113601,PhysRevA.97.042334}. Towards these goals, a crucial next step is to develop a numerically robust, cost-efficient recipe that combines the Bargmann invariant evaluation with functional integration and differentiation [as required by Eqs. (\ref{eq:QFI_waveform}) and (\ref{eq:fmWeveform})]. We leave these prospects for future study.

\newpage

\section{Derivation of the Generalized Replica Master Equation}
\label{sec:I}
In this appendix, we derive the generalized replica master equation (GRME), cf. Eq.~\eqref{eq:replicaME} of the main text, that serves as the central formula for the evaluation of the Bargmann invariants of the output field. 

First, let us introduce the continuous matrix-product operator (cMPO) representation of the joint density operator of the sensor and the waveguide,
\begin{equation}
\label{eq:rho_joint}
\rho_{\rm SW}(\theta,T)={\cal T}e^{ \int_0^Tdt[{\cal L}_0(\theta)+{\cal J}(\theta){\cal G}(t)+{\cal K}(\theta){\cal F}(t)]}\rho(0)\otimes\Xi(0).
\end{equation}
Equation~\eqref{eq:rho_joint} has a structure similar to the waveguide density operator $\Xi(\theta,T)$, cf. Eq.~\eqref{eq:rhoE} of the main text, albeit without tracing away the sensor. The waveguide density operator can therefore be expressed via Eq.~\eqref{eq:rho_joint} as $\Xi(\theta,T)={\rm tr}_{\rm S}[\rho_{\rm SW}(\theta,T)]$. In Eq.~\eqref{eq:rho_joint}, the temporal integral $\int_0^Tdt[\dots]$ should be interpreted as an It\^o integral, wherein the superoperators are defined the same as the main text---the sensor superoperator ${\cal L}_0$ via Eq.~\eqref{eq:no_jump_evol} of the main text, and ${\cal J,K}$ via ${\cal J}(\rho):=J\rho$, ${\cal K}(\rho):=\rho J^\dag$ with $J$ the sensor jump operator coupled with the waveguide; the waveguide superoperators ${\cal G}(t)(\Xi):=f^\dag(t)\Xi$ and ${\cal F}(t)(\Xi):=\Xi f(t)$, with $f(t)$ the quantum noise operator of the waveguide. 

The evolution equation of $\rho_{\rm SW}(\theta,T)$ can be derived by expanding the right hand side of Eq.~\eqref{eq:rho_joint} to ${\cal O}(dt)$, resulting in an It\^{o} quantum stochastic differential equation~\cite{carmichael1993open,RevModPhys.70.101,breuer2002theory,gardiner2004quantum,Wiseman,gardiner2015quantum}
\begin{equation}
\label{eq:SDE_Ito}
d{\rho}_{\rm SW}={\cal L}_0(\theta) \rho_{\rm SW}dt+d{F}^\dag(t)J(\theta)\rho_{\rm SW}+\rho_{\rm SW}J^\dag(\theta) d{F}(t),
\end{equation}
with the quantum noise increment of the waveguide defined as $d{F}(t):=f(t)dt$. As in the main text, we assume that the waveguide is initialized in the vacuum state (or coherent state, which can be transformed into the vacuum state via the Mollow transformation~\cite{PhysRevA.12.1919}). The quantum noise increment therefore satisfies the vacuum It\^{o} table~\cite{carmichael1993open,RevModPhys.70.101,breuer2002theory,gardiner2004quantum,Wiseman,gardiner2015quantum}
\begin{align}
\label{eq:Ito_table}
d{F}(t)d{F}^\dag(t)&=dt,\nonumber\\
d{F}^\dag(t)d{F}(t)&=d{F}(t)d{F}(t)=d{F}^\dag(t)d{F}^\dag(t)=0,
\end{align}
when acting on $\rho_{\rm SW}$ from the left. 

Now let us consider the Bargmann invariants of the waveguide $B(\Theta,T)$, cf. Eq.~\eqref{eq:Bargmann} of the main text. Using the fact that $\Xi(\theta,T)={\rm tr_{S}}[\rho_{\rm SW}(\theta,T)]$, we can express the waveguide Bargmann invariants as
\begin{align}
B(\Theta,T)=\,&{\rm tr}[\Xi(\theta_1,T)\Xi(\theta_2,T)\dots\Xi(\theta_{m+2},T)]\nonumber\\
\equiv\,&{\rm tr_{S_1,S_2,\dots,S_{m+2}}}{\rm tr}_{\rm W}\big[\rho_{\rm S_1W}(\theta_1,T)\rho_{\rm S_2W}(\theta_2,T)\nonumber\\
&\dots\rho_{\rm S_{m+2}W}(\theta_{m+2},T)\big]\nonumber\\
\equiv\,&{\rm tr}[\varrho(\Theta,T)],
\end{align}
where we define an operator acting on $m + 2$ replicas of the sensor,
\begin{equation}
\label{eq:varrho_def}
\varrho(\Theta,T):={\rm tr}_{\rm W}\big[\rho_{\rm S_1W}(\theta_1,T)\dots\rho_{\rm S_{m+2}W}(\theta_{m+2},T)\big]. 
\end{equation}
Since the sensor and the waveguide are initially unentangled, we have $\varrho(\Theta,T=0)=\otimes_{\alpha=1}^{m+2}\rho^{(\alpha)}(0)$, with $\rho(0)$ the sensor initial state. Moreover, the evolution of $\varrho(\Theta,T)$ can be derived straightforwardly via differentiating Eq.~\eqref{eq:varrho_def} and using Eq.~\eqref{eq:SDE_Ito} and the It\^o table Eq.~\eqref{eq:Ito_table},
\begin{align}
\label{eq:GRME_deriv}
d\varrho=&\sum_{\alpha=1}^{m+2}{\rm tr}_{\rm W}\big[\rho_{\rm S_{1}W}\dots {\cal L}_0(\theta_\alpha)\rho_{\rm S_{\alpha}W}\dots\rho_{\rm S_{m+2}W}\big]dt\nonumber\\
&+\sum_{\alpha=1}^{m+2}{\rm tr}_{\rm W}\big[\rho_{\rm S_{1}W}\dots \rho_{\rm S_{\alpha}W}J^{(\alpha)\dag}(\theta_\alpha)dF(t)\nonumber\\
&\qquad\quad\times dF^\dag(t) J^{(\alpha+1)}(\theta_{\alpha+1})\rho_{\rm S_{\alpha+1}W}\dots\rho_{\rm S_{m+2}W}\big]\nonumber\\
=&\sum_{\alpha=1}^{m+2}\left[{\cal L}_0(\theta_\alpha)\varrho+J^{(\alpha+1)}(\theta _{\alpha+1})\varrho J^{(\alpha)\dagger}(\theta_{\alpha})\right]dt
\end{align}
where we have used the shorthand denotation $\rho_{{\rm S}_\alpha {\rm W}}(\theta_\alpha,T)\equiv\rho_{{\rm S}_\alpha {\rm W}}$, and imposed periodic boundary condition $(m+3)=(1)$ on the replica index. Equation \eqref{eq:GRME_deriv} recovers the GRME~\eqref{eq:replicaME} of the main text.

\section{Quantum Fisher information and the Bargmann invariants}
\label{sec:appII}
In this appendix, we detail the derivation of the expansion formulas of the quantum Fisher information (QFI) for single-parameter estimation, Eqs.~\eqref{eq:In} and~\eqref{eq:fm} of the main text, in terms of derivatives of the Bargmann invariants. These expansion formulas are very general---in the following, we will present a general derivation for any parameter-encoded density operator $\Xi(\theta)$. For the special case considered in the main text, i.e., the sensing of an unknown parameter $\theta$ via continuous measurement of the output field of a quantum optical open system, $\Xi(\theta)=\Xi(\theta,T)$ corresponds to the density operator of the multi-photon state emitted by the open quantum sensor in the time window $[0,T]$. We will elaborate on this in Sec.~\ref{subsec:scenarios of continuous sensing} of this appendix. 

\subsection{Exponentially convergent series $I_n(\theta)$}
\label{appB1}
Consider a general parameter-dependent quantum state $\Xi(\theta)$. The QFI can be expressed as ${\rm tr}[\Xi(\theta) L^2]$, with $L$ the symmetric logarithmic derivative (SLD) operator defined via $\partial_{\theta}\Xi=(L\Xi+\Xi L)/2$. We denote the eigen-decomposition of the density operator as $\Xi(\theta)=\sum_{\lambda}{\lambda}\ket{\lambda}\bra{\lambda} $. The QFI of the state can hence be expressed in the eigenbasis as~\cite{Helstrom1976}
\begin{equation}
\label{eq:QFI}
I(\theta)=\frac{1}{2}\sum_{\lambda,\lambda^\prime} ({\lambda }+{\lambda '})|\langle \lambda |L |\lambda'\rangle|^2.
\end{equation}
Let us introduce a real, positive number $\Lambda$ that is larger than twice the largest eigenvalue of $\Xi(\theta)$, i.e., $\Lambda\geq 2 {\rm max}[\lambda]$. The precise value of $\Lambda$ is to be specified below. This allows us to expand Eq.~\eqref{eq:QFI} as
\begin{equation}
\label{eq:QFI_expansion}
I(\theta)=\frac{1}{2\Lambda}\sum_{\lambda,\lambda^\prime}\sum_{j=0}^{\infty}\left(1-\frac{\lambda+\lambda^\prime}{\Lambda}\right)^j(\lambda+\lambda^\prime)^2|\langle \lambda |L |\lambda'\rangle|^2,
\end{equation}
as can be verified straightforwardly. Let us define a convergent series $\{I_n(\theta)\mid n\in\mathbb{N}_0\}$ with
\begin{equation}
\label{eq:QFI_n}
I_n(\theta)=\frac{1}{2\Lambda}\sum_{\lambda,\lambda^\prime}\sum_{j=0}^{n}\left(1-\frac{\lambda+\lambda^\prime}{\Lambda}\right)^j(\lambda+\lambda^\prime)^2|\langle \lambda |L |\lambda'\rangle|^2.
\end{equation}
Apparently, $I_n(\theta)\leq I(\theta)$ and ${\rm lim}_{n\to \infty}I_n(\theta)=I(\theta)$. Moreover, the series converges exponentially fast with respect to $n$, as evident from evaluating the truncation error
\begin{align}
\label{eq:trunc_error}
I(\theta )-I_n(\theta )&= \sum _{ \lambda,\lambda'}\!\sum _{ j=n+1}^{\infty }\!\!\left(1-\frac{\lambda+\lambda^\prime}{\Lambda}\right)^j \!\frac{(\lambda+\lambda^\prime)^2}{2\Lambda}|\langle
\lambda |L |\lambda'\rangle|^2\nonumber\\
&\leq \sum _{ \lambda,\lambda^\prime}\frac{(\lambda+\lambda')^2}{2\Lambda}|\langle \lambda |L |\lambda'\rangle |^2\times\sum _{ j=n+1}^{\infty }\xi ^j\nonumber\\
&=C\times \xi^{n+1},
\end{align}
where we have defined 
\begin{equation}
\label{eq:xi_def}
\xi=1-\frac{2{\rm min}[\lambda]}{\Lambda},
\end{equation}
and $C=\sum _{ \lambda,\lambda^\prime}(\lambda+\lambda')^2|\langle \lambda |L |\lambda'\rangle |^2/[2\Lambda(1-\xi)]$ an $n$-independent coefficient. Therefore, the series $\{I_n(\theta)\mid n\in\mathbb{N}_0\}$ converges (at least) exponentially fast with $n$ towards $I(\theta)$. Additionally, the convergence speed is dependent on the choice of the parameter $\Lambda$---the smaller is $\Lambda$ (while satisfying the constraint $\Lambda\geq2{\rm max}[\lambda]$), the smaller is $\xi$ and hence the faster $I_n(\theta)$ converges towards $I(\theta)$. 

It is interesting to note that for a pure state, $\Xi(\theta)=\ket{\psi(\theta)}\bra{\psi(\theta)}$, $I_n(\theta)=I(\theta)$, $n=1,2,\dots$, i.e., each term of the series is identical to the QFI, as can be easily verified. 

\subsection{Relating $I_n(\theta)$ to the Bargmann invariants}
\label{appB2}
Our goal here is to derive Eqs.~\eqref{eq:In} and~\eqref{eq:fm} of the main text, i.e., to relate the approximate QFI $I_n(\theta)$ to the Bargmann invariants. The QFI $I_n(\theta)$ are defined in Eq.~\eqref{eq:QFI_n}, which can be converted to
\begin{align}
\label{eq:In_transform}
I_n(\theta )=&\sum_{ \lambda,\lambda'}\sum _{ j=0}^n\sum _{m=0}^j\binom{j}{m}(-1)^m\frac{({\lambda}+{\lambda '})^{m+2}}{2\Lambda^{m+1}}|\langle\lambda|L|\lambda' \rangle|^2\nonumber\\
=&2\sum _{ m=0}^n\sum _{ j=m}^n\binom{j}{m}\frac{(-1)^m}{\Lambda^{m+1}}\nonumber\\
&\times\sum _{ l=0}^m\binom{m}{l}{\rm tr}[\left(\partial _{\theta}\Xi \right)\Xi ^{m-l}\left(\partial
_{\theta }\Xi \right)\Xi ^l],
\end{align}
where we have used the definition of the SLD operator, and the identity $\sum _{ j=0}^n\sum _{ m=0}^j=\sum _{ m=0}^n\sum _{ j=m}^n$. Further exploiting the Hockey-Stick identity $\sum_{ j=m}^n\binom{j}{m}=\binom{n+1}{m+1}$, we arrive at
\begin{align}
I_n(\theta )&=\sum _{ m=0}^n\binom{n+1}{m+1}(-1)^mf_m(\theta)/\Lambda^{m+1},\label{In_app}\\
f_m(\theta )&=2\sum _{ l=0}^m\binom{m}{l}\text{tr}\left[\left(\partial _{\theta }\Xi \right)\Xi ^{m-l}\left(\partial _{\theta }\Xi \right)\Xi ^l\right].\label{fm_app}
\end{align}
Equation \eqref{In_app} recovers Eq.~\eqref{eq:In} of the main text. To confirm that Eq.~\eqref{fm_app} is equilavent to Eq.~\eqref{eq:fm}, we first carrying out the derivative $\partial^2_{\mu,\nu}$ in Eq.~(5b), thus converting it into 
\begin{align}
\label{eq:f_n_alt2}
f_m(\theta)&=\sum_{l=0}^m D_{l}^{m}\sum_{j=0}^l\sum_{k=0}^{m-l}{\rm tr}[\Xi^{j}(\partial_\theta\Xi)\Xi^{l-j+k}(\partial_\theta\Xi)\Xi^{m-l-k}]\nonumber\\
&\equiv\sum_{l,t=0}^m D_{l}^{m}\sum_{j=0}^l\sum_{k=0}^{m-l}{\rm tr}[(\partial_\theta\Xi)\Xi^t(\partial_\theta\Xi)\Xi^{m-t}]\delta_{t,l-j+k}.
\end{align}
It is straightforward to verify the following identity
\begin{align}
\sum_{j=0}^l\sum_{k=0}^{m-l}\delta_{t,l-j+k}=\left\{ \begin{array}{ll}
 t+1 &\,\,0\leq t\leq \ell_0,\\
 {\rm min}\{l,m-l\}+1 \quad\, &\ell_0\leq t \leq\ell_1,\\
 m-t+1  &\ell_1\leq t\leq m,
  \end{array} \right.
\end{align}
with $\ell_0:={\rm min}\{l,m-l\}$ and $\ell_1:={\rm max}\{l,m-l\}$. Using it, and the definition $D^{m}_l=2\binom{m}{l}-\binom{m}{l+1}-\binom{m}{l-1}$, it is straightforward to show that
\begin{align}
\label{eq:Didentity}
\sum_{l=0}^m D_{l}^{m}\sum_{j=0}^l\sum_{k=0}^{m-l}\delta_{t,l-j+k}=2\binom{m}{t},
\end{align}
hereby converting Eq.~\eqref{eq:f_n_alt2} to Eq.~\eqref{fm_app}.

Before we proceed, we note that expansion formulae spiritually similar to, albeit quantitatively different from Eqs.~\eqref{In_app} and \eqref{fm_app} [or equivalently, Eqs.~(5a) and (5b) of the main text] have been established in Ref.~\cite{PhysRevLett.127.260501} for the `quantum part' of the QFI $I_{\rm q}(\theta)$. This quantity is defined via splitting the full QFI Eq.~\eqref{eq:QFI} into the `classical' and `quantum' parts,
\begin{align}
I(\theta)=&\sum_{\lambda>0}\frac{1}{\lambda}\left(\frac{\partial \lambda}{\partial \theta}\right)^2+2\sum _{ \lambda +\lambda'>0}\frac{(\lambda-{\lambda'})^2}{{\lambda }+{\lambda '}}|\langle \lambda |\partial _{\theta}|\lambda'\rangle|^2\nonumber\\
=:&\,I_{\rm c}(\theta)+I_{\rm q}(\theta).
\end{align}
If the parameter is encoded via a unitary transformation $U(\theta)$, i.e., $\Xi(\theta)=U(\theta)\Xi U^\dag(\theta)$, the eigenvalues $\lambda$ are conserved hence $I_{\rm c}(\theta)=0$ and $I(\theta)=I_{\rm q}(\theta)$, as considered by Ref.~\cite{PhysRevLett.127.260501}. For general nonunitary encoding, such as the continuous sensing scenario studied in the main text, both $I_{\rm c}(\theta)$ and $I_{\rm q}(\theta)$ contribute to the QFI of the output field. For such scenarios, we should use the expansion formulae of the full QFI established in this work, Eqs.~\eqref{In_app} and \eqref{fm_app} [or equivalently, Eqs.~\eqref{eq:In} and \eqref{eq:fm} of the main text]. 

\subsection{Application to scenarios of continuous sensing}
\label{subsec:scenarios of continuous sensing}
\subsubsection{Choice of the free parameter $\Lambda$}
In the above, we have established for a general density operator $\Xi(\theta)$ the approximate formulas of its QFI, Eqs.~\eqref{In_app} and \eqref{fm_app}, with the parameter $\Lambda$ therein left unspecified. Now let us consider the sensing scenario studied in the main text, where $\Xi(\theta)=\Xi(\theta,T)$ is the density operator of the multi-photon state emitted by the open quantum sensor in the time window $[0,T]$, and our goal is to evaluate $I_n(\theta, T)$ efficiently through computing the Bargmann invariants of $\Xi(\theta,T)$. To this end, we further require that the evaluation of $\Lambda$ does not introduce extra computational cost besides the evaluation of the Bargmann invariants. Combining this consideration with the constraint that $\Lambda$ is larger than twice the largest eigenvalue of $\Xi(\theta,T)$, i.e., $\Lambda \geq 2{\rm max}[\lambda]$, the choices
\begin{equation}
\Lambda_q = 2[{\rm tr}(\Xi^q(\theta,T))]^{1/q}\equiv2\Big(\sum_\lambda \lambda^q\Big)^{1/q},
\label{eq:lambda_q}
\end{equation}
with $q=2,3,\dots$, are excellent as they are readily available from the Bargmann invariants, and they approach $2{\rm max}[\lambda]$ as $q\to\infty$. While a larger $q$ benefits faster convergence towards $I(\theta,T)$, in our numerical studies of continuous sensor models we find that the choice $q=2$ already suffices to ensure a reasonably fast convergence speed with respect to $n$. Hence, for the sack of concreteness, in the main text we have fixed $\Lambda=\Lambda_2 =2[{\rm tr}(\Xi^2(\theta,T))]^{1/2}$.

\subsubsection{Convergence speed considerations}
Finally, we comment on the convergence of our approach in continuous-sensing settings. With the choice of $\Lambda_q$ in Eq.~\eqref{eq:lambda_q}, the convergence of the truncated QFI series $\{I_n(\theta,T)\mid n\in\mathbb{N}_0\}$ is controlled by the parameter
\begin{equation}
\xi_q=1-\frac{{\rm min}[\lambda]}{(\sum_\lambda \lambda^q)^{1/q}}\gtrsim 1-\frac{{\rm min}[\lambda]}{{\rm max}[\lambda]},
\end{equation}
%Generically, the waveguide photons becomes increasingly entangled with the surrounding environments as $T$ increases, leading to an increasingly mixed density operator $\Xi(\theta,T)$.  As a result, $\xi_q$ typically increases with increasing $T$, resulting in a slower convergence speed for $\{I_n(\theta,T)\}$ towards $I(\theta,T)$. Therefore, to evaluate the QFI for very long $T$, it is not the optimal choice to directing evaluate $\{I_n(\theta,T)\}$ to high enough order. Instead, we can make use of the generic feature of the QFI growth: as long as the sensor steady state is reached, the QFI grows linearly with $T$. Hence, $I(\theta,T)$ for arbitrarily long $T$ can be extrapolated based on its intermediate-time behavior $I(\theta,T\simeq T_{\rm rel})$, with $T_{\rm rel}$ the relaxation time of the sensor to reach its steady state. Such intermediate-time QFI typically converges rather rapidly with our method.
as $I(\theta,T)-I_n(\theta,T)\lesssim {\cal O}(\xi_q^n)$. As $T$ increases, waveguide photons generally become more entangled with uncontrolled environmental degrees of freedom, so the reduced state $\Xi(\theta,T)$ becomes more mixed. This typically enlarges $\xi_q$, and therefore slows the convergence of $\{I_n(\theta,T)\}$ towards $I(\theta,T)$. As a result, for a very long interrogation time $T$, it may be inefficient to push the truncation order $n$ to very large values. Instead, one can exploit the generic late-time scaling: once the sensor has relaxed to its steady state, the QFI grows linearly in $T$. Thus, $I(\theta,T)$ at long $T$ can be obtained by estimating the linear growth rate from the intermediate-time regime $T\simeq T_{\rm rel}$, where $T_{\rm rel}$ is the sensor relaxation time. In this regime, the truncated series $\{I_n(\theta,T)\mid n\in\mathbb{N}_0\}$ typically converges much more rapidly.

\section{Extension to Waveform Estimation}
In this appendix, we extend the derivation in Apps.~\ref{sec:I} and \ref{sec:appII} to the scenario of waveform estimation, validating the discussion in the End Matter of the main text. 

First, we note that when the sensor dynamics depend on a waveform $\vartheta(t), t\in[0,T]$, the derivation in Eqs. (\ref{eq:rho_joint})--(\ref{eq:GRME_deriv}) remains valid under the substitution $\theta\to\vartheta(t)$. Consequently, the waveform-parameterized Bargmann invariants of the waveguide follow from the waveform-parameterized GRME, confirming Eqs.~\eqref{eq:BargmannWaveform} and \eqref{eq:replicaMEwaveform} in the End Matter.

Second, we relate the QCRB for waveform estimation to the waveform-parameterized Bargmann invariants. Consider an open quantum sensor subject to a waveform $\vartheta(t),t\in[0,T]$, and monitored continuously. The density operator of the waveguide photons emitted in the time window $[0,T]$ is a functional of the waveform, $\Xi\equiv\Xi[\vartheta(t)]$, which can be formally eigen-decomposed as $\Xi[\vartheta(t)]=\sum_\lambda \ket{\lambda}\bra{\lambda}$. In the basis $\{\ket{\lambda}\}$, the QFI matrix element for waveform estimation can be expressed as~\cite{Liu_2020}
\begin{align}
\label{eq:QFImat}
\tilde{{\cal I}}(t,t')=\frac{1}{2}\sum_{\lambda,\lambda'}(\lambda+\lambda'){\rm Re}[\langle \lambda| L_t |\lambda^\prime\rangle\langle\lambda^ \prime |L_{t'}|\lambda\rangle],
\end{align}
with Re denoting the real part, and the SLD operator $L_t$ defined as
\begin{align}
\frac{\delta\Xi[\vartheta(t)]}{\delta\vartheta(t)}=\frac{1}{2}(L_t\Xi+\Xi L_t).
\end{align}

Similarly to the derivation in App.~\ref{appB1}, we introduce $\Lambda =2\{{\rm tr}\Xi^2[\vartheta(t)]\}^{1/2}$ and expand Eq.~\eqref{eq:QFImat} as
\begin{align}
\label{eq:QFImat_expansion}
\tilde{\cal I}(t,t')=\frac{1}{2\Lambda}\sum_{\lambda,\lambda^\prime}\sum_{j=0}^{\infty}\left(1-\frac{\lambda+\lambda^\prime}{\Lambda}\right)^j(\lambda+\lambda^\prime)^2\nonumber\\
\times{\rm Re}[\langle \lambda| L_t |\lambda^\prime\rangle\langle\lambda^ \prime |L_{t'}|\lambda\rangle].
\end{align}
We can define the convergent series,
\begin{align}
\label{eq:InMat}
\tilde{\cal I}_n(t,t')=\frac{1}{2\Lambda}\sum_{\lambda,\lambda^\prime}\sum_{j=0}^{n}\left(1-\frac{\lambda+\lambda^\prime}{\Lambda}\right)^j(\lambda+\lambda^\prime)^2\nonumber\\
\times{\rm Re}[\langle \lambda| L_t |\lambda^\prime\rangle\langle\lambda^ \prime |L_{t'}|\lambda\rangle].
\end{align}
It can be readily verified, similarly to Eqs.~\eqref{eq:trunc_error} and~\eqref{eq:xi_def}, that the matrix elements $\tilde{\cal I}_n(t,t')$ converges (at least) exponentially fast with $n$ towards $\tilde{\cal I}(t,t')$.

Finally, by repeating the transformations described by Eqs.~(\ref{eq:In_transform}--\ref{eq:Didentity}) of Appendix~\ref{appB2}, we can re-express Eq.~\eqref{eq:InMat} as
\begin{align}
&\tilde{\cal I}_n(t,t')=\sum_{m=0}^n (-1)^m\binom{n+1}{m+1}\frac{f_m(t,t')}{\Lambda^{m+1}[{\vartheta(t)}]},\label{eq:InWeveform2}\\
&f_m(t,t')=\sum_{l=0}^m {D^{m}_l}\frac{\delta^2}{\delta\mu(t)\delta\nu(t')}{\rm tr}\!\left(\Xi[\mu]^{l+1}\Xi[{\nu}]^{m-l+1}\right)\!|_{{\mu}={\nu}={\vartheta}},
\end{align}
hence validating Eqs.~\eqref{eq:InWeveform} and \eqref{eq:fmWeveform} of the End Matter.

\section{Time-evolving Block Decimation Algorithm for Solving the Generalized Replica Master Equation}
Here we elaborate on the implementation of the time-evolving block decimation (TEBD) algorithm for solving the GRME, cf. Eq.~\eqref{eq:replicaME} of the main text. We provide strong numerical evidence and intuitive arguments showing that its solution satisfies the entanglement area law, which guarantees the efficiency and scalability of TEBD. The method extends to solving the waveform-parameterized GRME [cf. Eq.~\eqref{eq:replicaMEwaveform} in the End Matter of the main text].
%, and show that its solution satisfies the entanglement area law that guarantees the efficiency and scalability of TEBD. 

\subsection{Implementation of the TEBD}
While the GRME [cf. Eq.~\eqref{eq:replicaME} of the main text] is formulated under the periodic boundary condition (PBC), the TEBD algorithm is typically implemented with the open boundary condition (OBC). This choice is advantageous because, under OBC, matrix product states (MPSs) have a well-defined orthogonality center~\cite{RevModPhys.93.045003}, which facilitates bond dimension truncation during time evolution. In contrast, finite MPSs with PBC lack a well-defined orthogonality center, making the direct application of TEBD challenging~\cite{PAECKEL2019167998}.

\begin{figure}[t!]
\centering{} \includegraphics[width=0.48\textwidth]{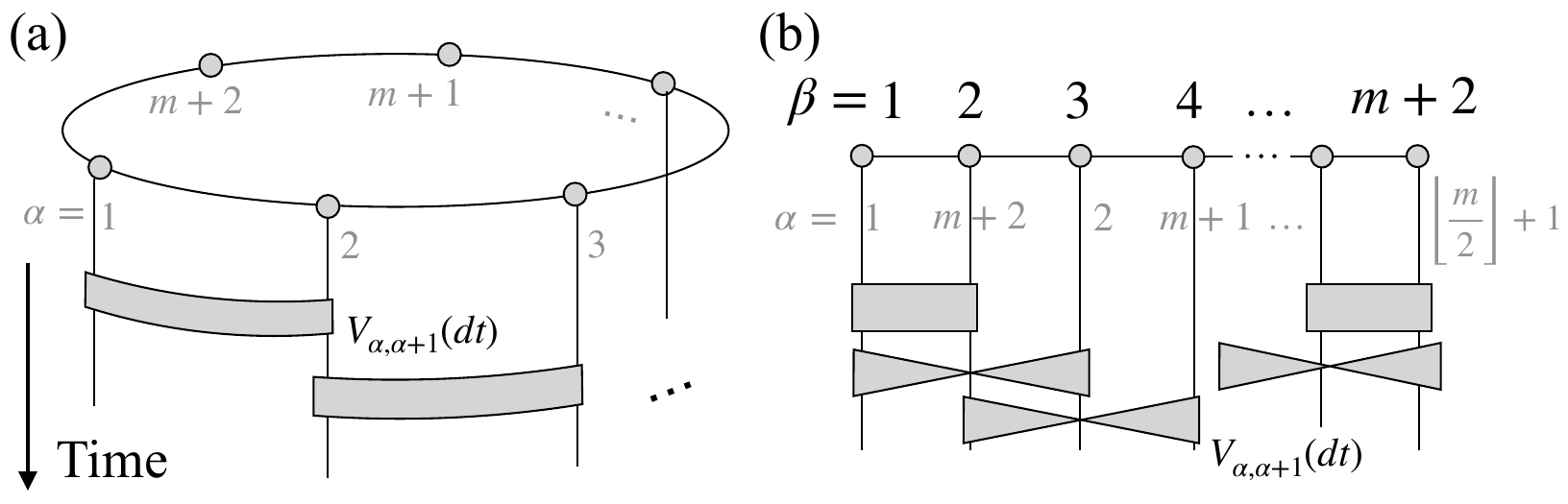} 
\caption{Folding the original periodic-boundary-condition replica chain (a) onto a one-dimensional lattice (b) with open boundary condition. As a result of folding, the nearest-neighbor two-replica gates $V_{\alpha,\alpha+1}(dt)$ in (a) are transformed to $m+2$ local gates in (b), involving $m$ next-nearest-neighbor gates (bow ties) across the lattice and two nearest-neighbor gates (rectangles) involving the open boundaries.}
\label{fig:fig_app_5} 
\end{figure}

To express $\varrho(\Theta,t)$ as a MPS with OBC, we adopt the folding approach as illustrated in Fig.~\ref{fig:fig_app_5}, where the original circular geometry (corresponding to PBC) of the replicas is folded onto a one-dimensional lattice with OBC. Denoting the site indices along the lattice by $\beta=1,2,\dots,m+2$, it is easy to verify that the site index $\beta$ is related to the original replica index $\alpha$ via
\begin{align}
\label{eq:alpha_beta}
\alpha &= \frac{\beta+1}{2},\,\qquad\qquad {\rm if} \,\,\, \beta={\rm odd},\nonumber\\
\alpha &= m+3-  \frac{\beta}{2},\qquad {\rm if} \,\,\, \beta={\rm even}.
\end{align}

On such an open-boundary lattice, the initial state of the GRME [cf. Eq.~\eqref{eq:replicaME} of the main text] is a product state of all sites, $\varrho(0)=\otimes_{\alpha=1}^{m+2} \rho^{(\alpha)}(0)\equiv \otimes_{\beta=1}^{m+2}\rho^{(\beta)}(0)$. The two-replica evolution operators $V_{\alpha,\alpha+1}(dt)$ in Eq.~\eqref{eq:TEBD} of the main text are represented by $m$ gates acting on \emph{next-nearest-neighbor} sites of the lattice, plus two \emph{nearest-neighbor} gates associated with the open boundaries, cf.~Fig.~\ref{fig:fig_app_5}(b). The two nearest-neighbor gates can be conveniently implemented, whereas the $m$ next-nearest-neighbor gates can be reduced to nearest-neighbor ones via swap operation between adjacent sites~\cite{PAECKEL2019167998}. The discrete-time propagation $\varrho(\Theta,t)\to\varrho(\Theta,t+dt)$ consists of implementation of all the $m+2$ Trotter gates, followed by truncation of the dimension of each bond $(\beta,\beta+1)$ to be below some threshold $\chi$. This results in a MPS representation of $\varrho(\Theta,t)$ on the open-boundary lattice,
\begin{equation}
\label{eq:MPS_replica}
\varrho(\Theta,t)=\!\sum_{\{s_\beta\}}\!A^{s_1}_{[1]}A^{s_2}_{[2]}\dots A^{s_{m+2}}_{[m+2]}s_{1}\otimes\dots \otimes s_{{m+2}},
\end{equation}
where $s_\beta$ (with $s\!=\!1,2,\dots,D^2$) labels the local basis of the single-replica Liouville space, and the matrices $A_{[\beta]}^{s_\beta}$ associated with site $\beta$ has dimension $\chi\times\chi$, with $\chi$ being the truncated bond dimension. 

\subsection{Entanglement area law of the solution of the generalized replica master equation}
As introduced in the main text, the dissipative nature of the GRME [cf. Eq.~\eqref{eq:replicaME} of the main text] leads to an area law of the bipartite entanglement entropy of $\varrho(\Theta,T)$ for all propagation time $T$, allowing for scaling the TEBD algorithm to large replica number and for long evolution time. Useful insights can be drawn from discretizing the GRME \eqref{eq:replicaME} in terms of the two-replica gates $V_{\alpha,\alpha+1}(dt)$, cf. Eq.~\eqref{eq:TEBD} of the main text, wherein the inter-replica correlation is solely established by the collective jump $\propto J^{(\alpha+1)}\varrho J^{(\alpha)\dag}$. Such jump \emph{annihilate excitations} of both replicas $\alpha$ and $\alpha+1$, and complete de-excitation of the replicas removes all entanglement across their bond. Thus, the collective jump serves a dual role—it both establishes inter-replica entanglement and simultaneously suppresses entanglement growth. Such feature is in stark contrast to the conventional unitary two-site gates, $\propto -i[H^{(\alpha,\alpha+1)},\varrho]$, which continuously drive entanglement growth. While a rigorous mathematical proof of the resultant entanglement area law is beyond the scope of this work, we present below compelling numerical evidence supporting its validity.

\begin{figure}[t!]
\centering{} \includegraphics[width=0.48\textwidth]{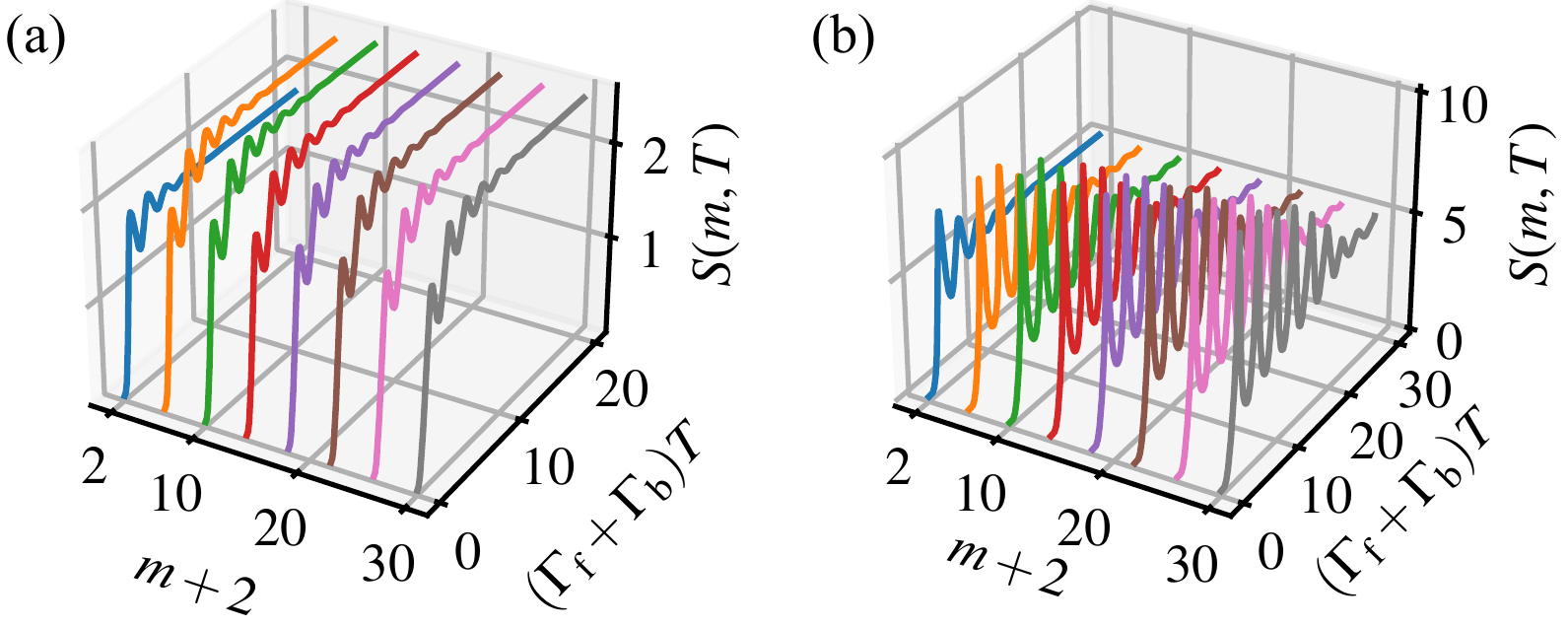} 
\caption{The maximum bipartite entanglement entropy $S(m,T)$ of the MPS Eq.~\eqref{eq:MPS_replica}, for various replica number $m+2$ and evolution time $T$, evaluated at the example of a $D$-level bidirectional emitter (see texts). We choose in (a) $D=2$ and in (b) $D=5$. The area law is clearly shown: $S(m,T)$ saturates with increasing replica number $m+2$ for all evolution time $T$. In the calculation we consider identical replicas with the parameters $\Delta=0,\Omega=5\Gamma_{\rm f},\Gamma_{\rm b}=0.5\Gamma_{\rm f}$, and we choose a sufficiently large truncation bond dimension $\chi=100$ to guarantee numerically exact calculation of the entanglement entropy. Note that this extreme bond dimension is not needed in an actual TEBD propagation---typically $\chi\lesssim 10$ suffices to calculate the Bargmann invariants accurately.}
\label{fig:fig_app_6} 
\end{figure}

To this end, we choose a sufficiently large truncation bond dimension $\chi$ for the MPS Eq.~\eqref{eq:MPS_replica} in our TEBD propagation, ensuring that the bipartite entanglement entropy $S_{\beta,\beta+1}(T)$ across each bond $(\beta,\beta+1)$ can be computed numerically exactly throughout the evolution. We define the maximum entropy among all bonds, $S(m,T)=\max_\beta S_{\beta,\beta+1}(T)$, which can in principle be dependent on the lattice size (i.e., the replica number) $m+2$ and the evolution time $T$. The area law, however, entails that $S(m,T)$ does not scale with the lattice size for all evolution time.

We verify such an area law behavior at the example of a bidirectional $D$-level emitter model. We denote the Fock basis for the $D$-level emitter as $\ket{d}$, $d=0,1,2,\dots,D-1$, and define the associated annihilation operator $a_D:=\sum_{d=0}^{D-2}\sqrt{d+1}\ket{d}\bra{d+1}$. The model is specified by the Hamiltonian $H=-\Delta a_D^\dag a_D+(\Omega a_D^\dag+{\rm h.c.})/2$ and  the jump operators $J_{\rm f(b)}=\sqrt{\Gamma_{\rm f(b)}}a_D$, with $\Gamma_{\rm f(b)}$ the emission rate into the forward, accessible (backward, inaccessible) channel. The case $D=2$ recovers the bidirectional two-level sensor studied in the main text. 

In Fig.~\ref{fig:fig_app_6}, we show the maximum entanglement entropy $S(m,T)$ as a function of the replica number $m+2$ and the evolution time $T$. $S(m,T)$ demonstrates a clear area law behavior---it saturates to a constant for increasing replica number $m+2$, and such saturation occurs for all evolution time $T$. Moreover, $S(m,T)$ increases with increasing sensor Hilbert space dimension $D$, as more excitations of the replicas can be produced hence more entanglement can be generated in the dissipative evolution according to the GRME \eqref{eq:replicaME} of the main text.

\section{Extension to scenarios involving non-Markovian noise}
Our method established in the main text provides a general framework for computing the optimal precision of continuous sensors subject to arbitrary Markovian noise. In this appendix, we show that the method can be extended to account for non-Markovian noise, via appropriate incorporation of recent advances~\cite{PhysRevLett.120.030402,doi:10.1142/S1230161222500196,PhysRevLett.132.100403} of the pseudomode approach~\cite{PhysRevA.50.3650,PhysRevA.55.2290} to describe non-Markovian dynamics. For simplicity, we consider sensing a single constant parameter $\theta$; the extension to waveform estimation is straightforward.

\subsection{Summary of the pseudomode approach}
\begin{figure}[t!]
\centering{} \includegraphics[width=0.48\textwidth]{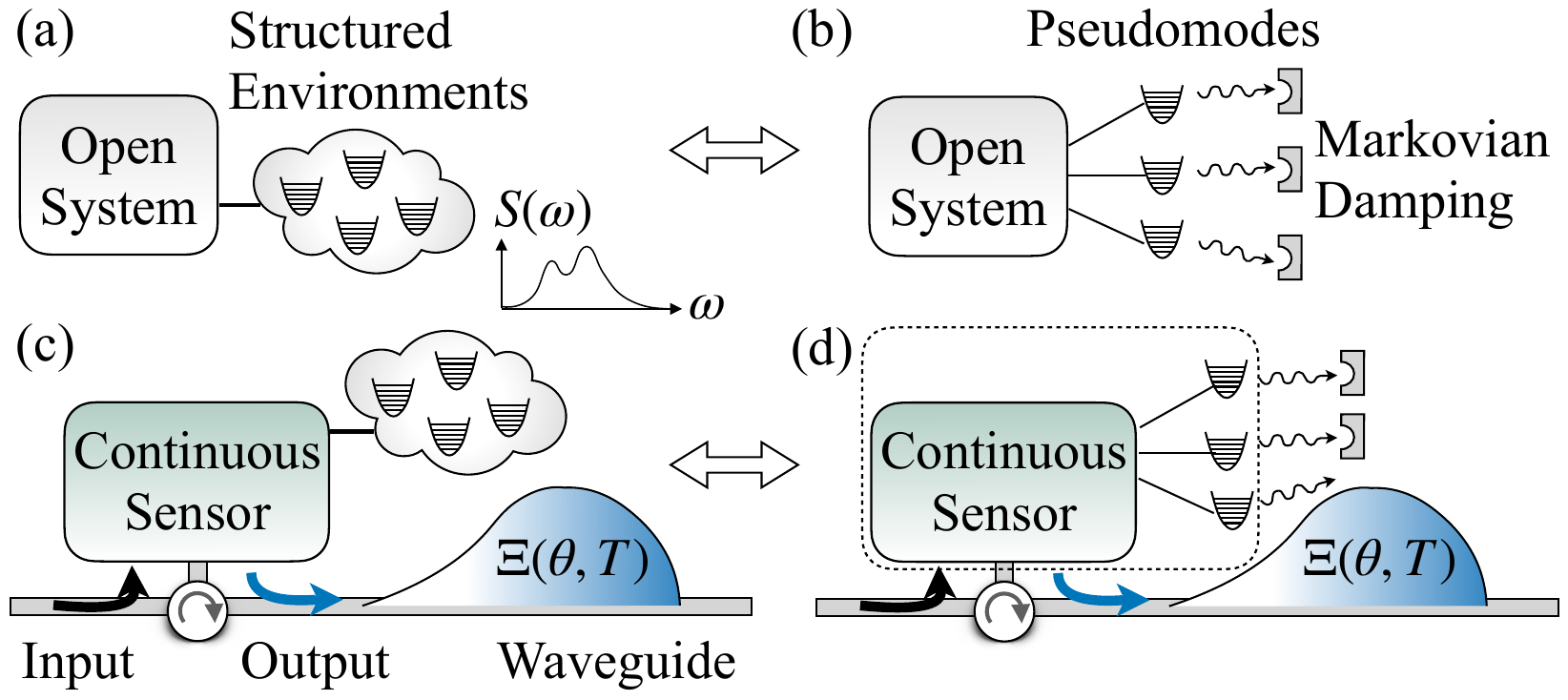} 
\caption{An open quantum system coupled to a non-Markovian environment, as shown in (a), can be mapped to an auxiliary configuration involving the original open system interacting with a small number of pseudomodes, each coupled to an independent Markovian environment, as shown in (b). Likewise, a sensor-waveguide interface coupled to a non-Markovian environment, as shown in (c), can be mapped to an auxiliary configuration where the composite system consisting of the sensor and the pseudomodes is interfaced with the waveguide, as shown in (d), and simultaneously subjected to Markovian noise. The configuration (d) can be efficiently solved using the method established in the present work, allowing us to evaluate the QFI of the waveguide photonic state $\Xi(\theta,T)$.}
\label{fig:fig_app_4} 
\end{figure}

The key idea of the pseudomode approach, as illustrated in Fig.~\ref{fig:fig_app_4}(a-b), is to map an original (potentially highly structured and non-Markovian) environment of the open system onto a simpler configuration, consisting of a small number of auxiliary bosonic modes---the pseudomodes---which in turn are coupled to independent Markovian environments. In this way, the reduced dynamics of the open system can be simulated by evolving the Lindblad master equation for the enlarged system consisting of the system itself and the pseudomodes, provided that the Hilbert-space of the enlarged system is of a reasonable size.

Specifically, let us consider a quantum system (S) interacting with a general bosonic environment (E), with the system-environment full Hamiltonian
\begin{equation}
\label{eq:SEoriginal}
H_{\rm SE} = H_{\rm S} + H_{\rm E} + A_{{\rm S}}\otimes F_{{\rm E}}.
\end{equation}
We assume that the system and environment are initialized in a factorized state $\rho_{\rm SE}(0)=\rho_{\rm S}(0)\otimes\rho_{\rm E}(0)$ and the initial environment state $\rho_{\rm E}(0)$ is a Gaussian state. To characterize the environment, we define the expectation values and two-time correlation functions with respect to its free evolution 
\begin{align}
\label{eq:corre_env}
\langle F_{{\rm E}}(t)\rangle&={\rm tr_E}[F_{\rm E} e^{-i H_{\rm E}t}\rho_{\rm E}(0)e^{i H_{\rm E}t}],\nonumber\\
C_{\rm E}(t,t')&={\rm tr_E}[e^{{i H_{\rm E}t}}F_{\rm E}e^{{-i H_{\rm E}(t-t')}}F_{\rm E}e^{{-i H_{\rm E}t'}}\rho_{\rm E}(0)].
\end{align}
We are interested in the reduced state of the system 
\begin{equation}
\rho_{\rm S}(t):={\rm tr}_{\rm E}[{\rm exp}(-i H_{\rm SE}t)\rho_{\rm SE}(0){\rm exp}(i H_{\rm SE}t)].
\end{equation}

Now we introduce an auxiliary configuration consisting of the open system interacting with $R$ pseudomodes via the Hamiltonian
\begin{equation}
H_{\rm SP}=H_{\rm S} + H_{\rm P} + A_{\rm S}\otimes G_{\rm P},
\end{equation}
with $H_{\rm P} $ the free Hamiltonian of the pseudomodes, and each pseudomode (labelled as $r=1,2,\dots,R$) is coupled to its own Markovian environment via the jump operator $M_r$. The evolution of the auxiliary configuration is hence described by the LME~\cite{PhysRevLett.120.030402,doi:10.1142/S1230161222500196,PhysRevLett.132.100403}
\begin{equation}
\label{eq:LMEpseudo}
\dot{\rho}_{\rm SP}=-i[H_{\rm SP},\rho_{\rm SP}]+\sum_{r=1}^R{\cal D}[M_{r}]\rho_{\rm SP}.
\end{equation}
We define the reduced state of the system in the auxiliary configuration as 
\begin{equation}
\tilde{\rho}_{\rm S}(t):={\rm tr}_{\rm P}[\rho_{\rm SP}(t)],
\end{equation}
To characterize the auxiliary configuration, we introduce the expectation values and two-time correlation functions with respect to the free evolution of the pseudomodes,
\begin{align}
\langle G_{\rm P}(t)\rangle&={\rm tr_P}[G_{\rm P}e^{{\cal L}_{\rm P}t}\rho_{\rm P}(0)],\nonumber\\
C_{\rm P}(t,t')&={\rm tr_P}[G_{\rm P}e^{{\cal L}_{\rm P}(t-t')}G_{\rm P}e^{{\cal L}_{\rm P}t'}\rho_{\rm P}(0)],
\end{align}
where ${\cal L}_{\rm P}\rho:=-i[H_{\rm P},\rho]+\sum_{r=1}^R{\cal D}[M_{r}]\rho$. Remarkably, Ref.~\cite{PhysRevLett.120.030402,doi:10.1142/S1230161222500196} proved an exact equivalence between the two configurations, provided that the expectation values and two-time correlation functions are the same, i.e.,
\begin{align}
\label{pseudomodeCondition}
\langle F_{{\rm E}}(t)\rangle&=\langle G_{\rm P}(t)\rangle,\quad\forall t>t'>0,\nonumber\\
C_{\rm E}(t,t')&=C_{\rm P}(t,t'),\nonumber\\
\Longrightarrow\rho_{\rm S}(t)&=\tilde{\rho}_{S}(t),\quad\forall t>0.
\end{align}

The pseudomode approach allows us to reproduce the open system dynamics subject to a general system-environment coupling Eq.~\eqref{eq:SEoriginal}, by solving a LME \eqref{eq:LMEpseudo} for the system and the pseudomodes. In Eq.~\eqref{eq:LMEpseudo}, the operators associated with the pseudomodes, i.e., $H_{\rm P}$, $G_{\rm P}$  and $M_r$, $r=1,2,\dots R$, are adjustable parameters and are required to reproduce the environmental expectation values and two-time correlation functions, cf. Eq.~\eqref{pseudomodeCondition}. 

\subsection{Incorporation of the pseudomode approach into the description of noisy continuous sensors}
In a general scenario of quantum sensing via continuous monitoring, as shown in Fig.~\ref{fig:fig_app_4}(c-d), an open sensor (S) is coupled with a waveguide (W), and both may interact with (potentially non-Markovian) environments (Es). For simplicity, let us consider the situation that the sensor is coupled with one such environment; extension to more than one environments is straightforward. 
The global Hamiltonian of the sensor, waveguide and environment can be written in the Schr\"odinger picture as
\begin{equation}
\label{eq:schHglob}
{\cal H}_{\rm Glob} = {\cal H}_{\rm S} +{\cal H}_{\rm W} +{\cal H}_{\rm SW} + {\cal H}_{\rm E}+ A_{{\rm S}}\otimes F_{{\rm E}},
\end{equation}
in which ${\cal H}_{\rm S}$, ${\cal H}_{\rm W}=\int d\omega \omega f^\dag(\omega)f(\omega)$ and ${\cal H}_{\rm E}=\int d\omega \omega b^\dag(\omega)b(\omega)$ are respectively the free Hamiltonian of the sensor, the waveguide and the environment, and 
\begin{equation}
{\cal H}_{\rm SW}=\frac{i}{\sqrt{2\pi}}\int^{\bar{\omega}+{\cal B}}_{\bar{\omega}-{\cal B}}d\omega[J_{\rm S}f^\dag(\omega)-J_{\rm S}^\dag f(\omega)]
\end{equation}
is the sensor-waveguide coupling. In ${\cal H}_{\rm SW}$, $J_{\rm S}$ is the system transition (jump) operator that interacts with photons in a bandwidth $\cal B$ around some mean optical frequency $\bar{\omega}$, and counterrotating terms have been neglected. Finally, the sensor-environment coupling is left unspecified and is described by the general term $A_{{\rm S}}\otimes F_{{\rm E}}$.

We can transform away the optical frequencies in Eq.~\eqref{eq:schHglob} by moving into a suitable interaction picture. If $A_{\rm S}$ involves the sensor transition operator $J_{\rm S}$ (and hence is fast oscillating at frequency $\bar{\omega}$), the interaction picture is with respect to ${\cal H}_{\rm E}+\bar{\omega} [J_{\rm S}^\dag J_{\rm S}+\int d\omega b^\dag({\omega})b(\omega)]$; otherwise it is with respect to ${\cal H}_{\rm E}+\bar{\omega} J_{\rm S}^\dag J_{\rm S}$. In this picture, the global Hamiltonian of the sensor, waveguide and environment becomes
\begin{equation}
\label{eq:intHglob}
H_{\rm Glob} = H_{\rm} +H_{\rm SW} + H_{\rm E}+ A_{{\rm S}}\otimes F_{{\rm E}},
\end{equation}
where $H$ describes the (slow) free evolution of the sensor, and $H_{\rm SW}=i[Jf^{\dag}(t)-{\rm h.c.}]$ is the sensor-waveguide coupling, with $f(t)$ the quantum noise operator for the waveguide modes,
\begin{equation}
f(t)=\frac{1}{\sqrt{2\pi}}\int_{{\cal B}}d\omega f(\omega)e^{-i(\omega-\bar{\omega})t}.
\end{equation}
In this picture, the free Hamiltonian of the environment reads $H_{\rm E}=\int d\omega \omega b^\dag(\omega)b(\omega)$ [or $H_{\rm E}=\int d\omega(\omega-\bar{\omega}) b^\dag(\omega)b(\omega)$], if the environment couples with the optical transition operator of the sensor [or if not]. The interaction-picture Hamiltonian Eq.~\eqref{eq:intHglob} describes the \emph{slow} dynamics of the sensor-waveguide interface, which further couples with a general bosonic environment. 

Crucially, the pseudomode approach holds even for open systems with an infinite-dimensional Hilbert space (e.g., a quantum field). This allows us to formally regard the sensor and the waveguide as a composite open system, and to apply the pseudomode treatment presented above to Eq.~\eqref{eq:intHglob}. It replaces the environment by a few pseudomodes coupled with the open sensor, as shown in Fig.~\ref{fig:fig_app_4}(d). Consequently, the reduced density operator of the waveguide can be evaluated via $\Xi(t)={\rm tr_{S,P}}[\rho_{\rm SWP}(t)]$, with $\rho_{\rm SWP}$ satisfying a LME
\begin{align}
\label{eq:pseudo_waveguide}
\dot{\rho}_{\rm SWP}&=-i[H_{\rm SWP},\rho_{\rm SWP}]+\sum_{r=1}^R{\cal D}[M_{r}]\rho_{\rm SWP},\nonumber\\
H_{\rm SWP}&=H+i[Jf^\dag(t)-{\rm h.c.}]+H_{\rm P}+A_{\rm S}\otimes G_{\rm P}.
\end{align}
In Eq.~\eqref{eq:pseudo_waveguide}, the operators of the pseudomodes, i.e., $H_{\rm P}$, $G_{\rm P}$ and $M_r$, $r=1,2,\dots,R$, are adjustable parameters and are required to reproduce the expectation values and two-time correlation functions of the original environment. Although perfectly matching these quantities typically requires an infinite number of pseudomodes, accurate approximations can be achieved with only a small number \cite{PhysRevLett.132.100403}. Moreover, the error in system observables can be rigorously upper-bounded in terms of the error in matching the the expectation values and two-time correlation functions~\cite{PhysRevLett.118.100401}.

To solve Eq.~\eqref{eq:pseudo_waveguide}, we first note that it involves the quantum noise operators $f(t)$ and $f^\dag(t)$ of the waveguide and, in the language of quantum stochastic calculus, should be interpreted as a quantum stochastic equation of the \emph{Stratonovich} form~\cite{carmichael1993open,RevModPhys.70.101,breuer2002theory,gardiner2004quantum,Wiseman}. We can convert Eq.~\eqref{eq:pseudo_waveguide} into the It\^o form following standard procedures~\cite{carmichael1993open,RevModPhys.70.101,breuer2002theory,gardiner2004quantum,Wiseman,gardiner2015quantum}, and thereby writing down its solution as a cMPO
\begin{align}
\label{eq:rhoSWP}
\rho_{\rm SWP}(T)={\cal T}e^{ \int_0^Tdt[{\cal L}_0+{\cal J}{\cal G}(t)+{\cal K}{\cal F}(t)]}\rho_{\rm SWP}(0),
\end{align}
where the superoperators are defined the same as the main text---the sensor superoperators ${\cal J,K}$ via ${\cal J}(\rho):=J\rho$, ${\cal K}(\rho):=\rho J^\dag$; the waveguide superoperators ${\cal G}(t)(\Xi):=f^\dag(t)\Xi$ and ${\cal F}(t)(\Xi):=\Xi f(t)$. Finally, ${\cal L}_0$ is a superoperator for the composite system consisting of the sensor and the pseudomodes,
\begin{align}
\label{eq:no_jump_evol_pseudo}
{\cal L}_{\rm 0}\rho:=&-i[H+H_{\rm P}+A_{\rm S}\otimes G_{\rm P},\rho]\nonumber\\
&-\frac{1}{2}\{J^\dag J,\rho\}
+\sum_{r=1}^R{\cal D}[M_r]\rho.
\end{align}

The above analysis shows that, with the adoption of the pseudomode description of non-Markovian noise, the waveguide density operator $\Xi(t)$ can be represented in a cMPO form similar to the Markovian scenario, provided that the local tensors of the cMPO include dynamical operators of both the sensor and the pseudomodes. As a result, the Bargmann invariants of the waveguide can again be evaluated via the GRME (6) of the main text, with ${\cal L}_0(\theta)$ therein replaced by Eq.~\eqref{eq:no_jump_evol_pseudo}. This provides us with a remarkably simple and convenient method to evaluate the precision limit of continuous sensors subject to non-Markovian noise, as summarized by the recipe below:

({$\bold 1$})~Given the structured, non-Markovian environment, construct the appropriate pseudomodes to reproduce (approximately) its expectation values and two-time correlation functions in Eq.~\eqref{eq:corre_env}.

({$\bold 2$})~Evaluate the Bargmann invariants of the waveguide via propagating the GRME [cf. Eq.~(6) of the main text] for the composite system consisting of the open sensor and the pseudomodes.

({$\bold 3$})~Evaluate the QFI of the waveguide from the Bargmann invariants via Eqs.~\eqref{eq:In} and~\eqref{eq:fm} of the main text.

\subsection{Illustration of the method}
Let us illustrate the method at the example of a driven-dissipative two-level sensor interfaced unidirectionally with a waveguide, and simultaneous subjected to non-Markovian dephasing, cf. Fig.~\ref{fig:fig3}(c) of the main text. The global Hamiltonian of the sensor, waveguide and environment in the interaction picture is given by Eq.~\eqref{eq:intHglob}, wherein
\begin{equation}
H=-\Delta \sigma_{ee}+(\Omega \sigma_{eg}+{\rm h.c.})/2, 
\end{equation}
with $\Delta$($\Omega$) the laser detuning (Rabi frequency) and $\sigma_{ij}:=\ket{i}\bra{j}$. The sensor couples with the waveguide via the jump operator $J=\sqrt{\Gamma}\sigma_{ge}$, with $\Gamma$ the emission rate. The sensor-environment coupling is
\begin{equation}
A_{{\rm S}}\otimes F_{{\rm E}}=\sigma_z\otimes\int d\omega[g(\omega) b(\omega)+{\rm h.c.}].
\end{equation}
For simplicity, we assume that both the waveguide and the environment are initialized in the vacuum state, $\rho_{\rm W,E}(0)=\ket{{\rm vac}}_{\rm W,E}\bra{{\rm vac}}$, and the sensor is initially in the ground state, $\rho_{\rm S}(0)=\ket{{g}}\bra{g}$. As a result, the environmental expectation values and two-time correlation functions in Eq.~\eqref{eq:corre_env} can be evaluated conveniently as 
$\langle F_{{\rm E}}(t)\rangle=0$, and
\begin{align}
C_{\rm E}(t+\tau,t)=\int_{-\infty}^{\infty}d\omega S_{\rm E}(\omega)e^{-i\omega \tau},
\end{align}
with $S_{\rm E}(\omega)=|g(\omega)|^2$ the environmental noise spectrum (or correlation spectrum). 

We construct the pseudomodes as $R$ independently damped oscillators initialized in their ground state, with $H_{\rm P}=\sum_{r=1}^R \nu_r c^\dag_r c_r$, and $M_r=\sqrt{\gamma_r}c_r$ been the jump operator of the $r$-th mode. Moreover, we choose the sensor-oscillator coupling as
\begin{align}
A_{{\rm S}}\otimes G_{{\rm P}}=\sigma_z\otimes\sum_{r=1}^R \left(g_r c_r+{\rm h.c.}\right),
\end{align}
with $g_r$ specifying the interaction strength between the sensor and the $r$-th oscillator. In the pseudomode description, the expectation values and two-time correlation functions hence follow as $\langle G_{\rm P}(t)\rangle=0$, and 
\begin{align}
C_{\rm P}(t+\tau,t)&=\int_{-\infty}^{\infty}d\omega S_{\rm P}(\omega)e^{-i\omega \tau},\nonumber\\
S_{\rm P}(\omega)&=\sum_{r=1}^{R}\frac{|g_r|^2\gamma_r}{\gamma_r^2+(\omega-\nu_r)^2}.
\end{align}
The requirement that $C_{\rm P}(t+\tau,t)=C_{\rm E}(t+\tau,t)$ therefore reduces to fitting the known environmental spectrum $S_{\rm E}(\omega)$ as a sum of a few Lorentzian profiles corresponding to each pseudomodes. 

In the numerical example shown in Fig.~\ref{fig:fig3}(c-d) of the main text, we consider a non-Markovian environment with an exact Lorentzian noise spectrum centered at zero frequency, $S_{\rm E}(\omega)=\gamma^2/(\omega^2+\gamma^2)$. Note that this noise spectrum is chosen such that $S_{\rm E}(0)$, which characterizes the system dephasing rate in Markovian limit~\cite{breuer2002theory}, is independent of $\gamma$. Such an environmental noise spectrum can be reproduced with a single damped oscillator, i.e., $R=1$, with its oscillation frequency $\nu=0$, its jump operator $M=\sqrt{\gamma}c$ and its sensor-oscillator coupling $g=\sqrt{\gamma}$. In the numerical simulation, we choose the sensor parameters $\Delta=0,\Omega=\Gamma$, and leave the environmental spectral width $\gamma$ adjustable. By solving the GRME [cf. Eq.~\eqref{eq:replicaME} of the main text] for the composite system consisting of the sensor and the pseudomode, we arrive at Fig.~\ref{fig:fig3}(d) of the main text.

\bibliography{references}

\end{document}